\documentclass[onecolumn]{autart}    

\usepackage[dvips]{epsfig}    
\usepackage{amsmath,mathtools,amssymb}
\usepackage{color}
\usepackage{algorithm}
\usepackage{algpseudocode}
\usepackage{comment}

\newtheorem{theorem}{Theorem}
\newtheorem{lemma}[theorem]{Lemma}

\newtheorem{proposition}[theorem]{Proposition}

\newtheorem{ass}{Assumption}
\newtheorem{definition}{Definition}

\newcommand{\R}{\mathbb{R}}
\newcommand{\N}{\mathbb{N}}

\newcommand{\step}[1]{\textit{Step #1:}}

\newcommand\D{\mathrm d}

\newcommand\norm[1]{\ensuremath{\lVert#1\rVert}}

\DeclareMathOperator\diag{diag}

\begin{document}

\begin{frontmatter}

\title{Recurrent neural network-based robust control systems with regional properties and application to MPC design\thanksref{footnoteinfo}} 

\thanks[footnoteinfo]{This paper was not presented at any IFAC 
meeting. Corresponding author D.~Ravasio.}

\author[Milano,Milano2]{Daniele Ravasio}\ead{daniele.ravasio@polimi.it, daniele.ravasio@stiima.cnr.it},    
\author[Milano]{Alessio La Bella}\ead{alessio.labella@polimi.it},
\author[Milano]{Marcello Farina}\ead{marcello.farina@polimi.it},               
\author[Milano2]{Andrea Ballarino}\ead{andrea.ballarino@stiima.cnr.it}  

\address[Milano]{Dipartimento di Elettronica, Informazione e Bioingegneria, Politecnico di Milano, 20133, Milano, Italy}  
\address[Milano2]{Istituto di Sistemi e Tecnologie Industriali Intelligenti per il Manifatturiero Avanzato, Consiglio Nazionale delle Ricerche, 20133, Milano, Italy}             


\begin{abstract}            
This paper investigates the design of output-feedback schemes for systems described by a class of recurrent neural networks. We propose a procedure based on linear matrix inequalities for designing an observer and a static state-feedback controller. The algorithm leverages global and regional incremental input-to-state stability ($\delta$ISS) and enables the tracking of constant setpoints, ensuring robustness to disturbances and state estimation uncertainty. To address the potential limitations of regional $\delta$ISS, we introduce an alternative scheme in which the static law is replaced with a tube-based nonlinear model predictive controller (NMPC) that exploits regional $\delta$ISS properties. We show that these conditions enable the formulation of a robust NMPC law with guarantees of convergence and recursive feasibility, leading to an enlarged region of attraction. Theoretical results are validated through numerical simulations on the pH-neutralisation process benchmark.
\end{abstract}

\end{frontmatter}

\section{Introduction}
In the last decades, the increased availability of plant measurements has led to growing interest in data-based control approaches \cite{tang2022data}. 
In particular, within the framework of indirect data-based control, recurrent neural network (RNN) models have gained increasing attention for their ability to capture nonlinear dynamics and are being investigated as system models for model-based control design~\cite{bonassi2022recurrent}.\\
Within this setting, a widespread approach to simplify the control design consists in constraining the RNN training procedure to ensure that the identified model satisfies specific stability properties (such as input-to-state stability (ISS) \cite{jiang2001input} or incremental ISS ($\delta$ISS) \cite{bayer2013discrete}). This approach has been applied, for instance, to the development of nonlinear model predictive control (NMPC) algorithms \cite{terzi2021learning,bonassi2021nonlinear,bonassi2024nonlinear,schimperna2024robust}, internal model control (IMC) architectures \cite{furieri2024learning,bonassi2022recurrent_imc}, and state observers \cite{bonassi2024nonlinear,schimperna2024robust} with proven theoretical guarantees. 
However, enforcing sufficient and possibly conservative conditions during model identification can restrict the set of admissible parameter values, possibly excluding parameters that accurately describe the system dynamics. Moreover, this approach is applicable only to systems characterised by the desired open-loop properties.\\
Therefore, it is important to study new conditions to impart these properties to the RNN-based control system. 
Note that, closed-loop $\delta$ISS is crucial for the development of robust controllers. For example, it enables the definition of robust positively invariant (RPI) sets \cite{bayer2013discrete}, which are essential in the design of tube-based NMPC schemes \cite{mayne2000constrained} and which, in the output-feedback framework, are essential to develop integrated design procedures that explicitly account for potential state estimation uncertainty \cite{shamma1998output,mayne2006robust}.\\
Only a few studies investigate the design of RNN-based output-feedback control schemes \cite{seel2021neural,d2023incremental,ravasio2024lmi,la2025regional}. However, most of these works focus on equilibrium stability, while only \cite{d2023incremental,ravasio2024lmi} investigate design conditions for enforcing closed-loop $\delta$ISS. 
\\
The main limitation of the cited approaches is that they aim to enforce global properties to the control system. However, global $\delta$ISS is a strong property that may not always be admissible, for example, due to the nonlinearities of the RNN model. Nevertheless, global stability notions are often unnecessary in practice, as physical systems typically operate within bounded regions due to saturation limits. To address this limitation, local conditions for the design of stabilising controllers should be investigated.\\
For the case of linear systems subject to saturation limits, \cite{massimetti2009linear} derived LMI conditions for the design of anti-windup control systems with regional stability. The key idea is to represent the saturation nonlinearity through a generalised sector condition. Importantly, generalised sectors encompass classical global sector conditions, allowing a local characterisation of the nonlinearity and introducing additional degrees of freedom in the formulation, which results in less conservative solutions. Leveraging the structural similarities between saturation functions and the sigmoidal activation functions, this approach has been extended in \cite{la2025regional} to the control of a class of RNN models. 
However, these studies primarily focus on equilibrium stability, and, to the best of the authors' knowledge, no prior work has addressed the problem of 
enforcing regional $\delta$ISS to the RNN-based control system.\\
In this work, we consider the control of a plant described by an RNN model, subject to disturbances and measurement noise. In particular, we focus on the class of RNNs introduced in [15].  The main contributions of this work are as follows:\\ 
\emph{(i)}	We derive a novel generalised incremental sector condition for a class of activation functions commonly used in RNN models. We exploit this description to derive LMI-based sufficient conditions for imparting closed-loop global or regional $\delta$ISS to the RNN. Notably, the resulting conditions can always be satisfied in a sufficiently small region under very mild assumptions.\\
\emph{(ii)} Based on these conditions, we design an output-feedback scheme, integrating a nonlinear state observer and a static state-feedback controller, that guarantees robust tracking of constant setpoints in the presence of disturbances and state estimation uncertainty.\\ 
\emph{(iii)} We address the potentially limited feasibility region of the static controller by showing that it can be used as an auxiliary law in the design of a theoretically sound tube-based NMPC, thereby enlarging the setpoint region of attraction.

The rest of the paper is organised as follows. Section~\ref{sec:problem_statement} introduces the RNN model, the control objective and the incremental sector condition. The design conditions for the observer, static state-feedback law and NMPC law are presented in Sections~\ref{sec:observer}, \ref{sec:local_dISS}, and \ref{sec:MPC}, respectively. Simulation results are provided in Section~\ref{sec:simulations}. Conclusions are drawn in Section~\ref{sec:conclusions}. The proofs of the main results are reported in the Appendix.
%
%
 \subsection*{Preliminaries} Given a vector \(v \in \mathbb{R}^n\), \(v^\top\) represents its transpose and \(v_i\) its \(i\)-th entry. Given a matrix \(M \in \mathbb{R}^{n\times n}\), \(m_{ij}\) denotes its entry in position \((i,j)\) and $M_i$ its $i$-th row. We denote the identity function as \(\text{id}(\cdot)\), the set-interior function as \(\mathcal{INT}(\cdot)\), and the \(n \times n\) identity matrix as \(I_n\). Given $n$ matrices $M^{(1)},\dots,M^{(n)}$, we denote by $\diag(M^{(1)},\dots,M^{(n)})$ the block-diagonal matrix with $M^{(1)},\dots,M^{(n)}$ on its main diagonal blocks. The limit superior is expressed as $\overline{\lim}$ \cite{jiang2001input}. The sequence  $\{u(k), \dots, u(k+N)\}$ is denoted as $u([k:k+N])$. 
Let $\R_{\geq 0}$ (resp., $\mathbb Z_{\geq 0}$) be the set of positive real (resp., positive integer) numbers including $0$ and $\R_{+}\coloneqq\R_{\geq 0}\backslash \{0\}$. Given a set $\mathcal{X}$, we denote the cartesian product of $\mathcal X$ taken $n$ times as $\mathcal{X}^n$.
Moreover, denote the set of positive definite real matrices as $\mathbb{S}_+^n$, and the set of diagonal positive definite real matrices as
$\mathbb{D}_+^n$.
For $Q\in\mathbb S_+^n$, let $\mathcal{E}(Q)=\{v\in\R^n:v^\top Qv\leq1\}$ denote the corresponding ellipsoidal set .
Moreover, \(\mathcal{B}_{\epsilon}^{(n)}(0)\) denotes a ball of radius \(\epsilon\), centered at \(0\), in \(\mathbb{R}^n\).
A continuous function $\alpha:\R_{\geq0}\to\R_{\geq0}$ is a class $\mathcal K$-function if $\alpha(s)>0$ for all $s>0$, it is strictly increasing and $\alpha(0)=0$. Also, a continuous function $\alpha:\R_{\geq0}\to\R_{\geq0}$ is a class $\mathcal K_\infty$-function if it is a class $\mathcal K$-function and $\alpha(s)\to\infty$ as $s\to\infty$. Finally, a continuous function $\beta:\R_{\geq0}\times\mathbb Z_{\geq0}\to\R_{\geq0}$ is a class $\mathcal{KL}$-function if $\beta(s,k)$ is a class $\mathcal K$-function with respect to $s$ for all $k$, it is strictly decreasing in $k$ for all $s>0$, and $\beta(s,k)\to0$ as $k\to\infty$ for all $s>0$. The pair $(A,B)$ (resp., $(A,C)$) is said to be stabilizable (resp., detectable) if there exists a matrix 
$K\in\mathbb{R}^{m\times n}$ (resp., $L\in\mathbb{R}^{n\times p}$) such that 
$A+BK$ (resp., $A-LC$) is stable, i.e., all its roots are inside the unit circle \cite{glad2018control}. In the following, we refer to the notions of $\delta$ISS, dissipation-form $\delta$ISS Lyapunov functions, and ISS observer error dynamics reported in Appendix~\ref{appendix_B}. 
%
%
%
%
\section{Problem statement}\label{sec:problem_statement}
\subsection{The plant model}
In this paper, we address the problem of controlling a nonlinear plant whose dynamics is described by the following RNN model class~\cite{la2025regional},
\begin{equation}\label{eq:plant_dynamics}
\begin{cases}
x(k+1)=A_xx(k)+B_uu(k)+D_ww(k)+B_\sigma\sigma(v(k))\\
v(k)=\tilde{A}x(k)+\tilde{B}u(k)+\tilde Dw(k)\\
y(k)=Cx(k)+\eta(k)
\end{cases}\ 
\end{equation}
where $x\in \R^n$ denotes the state vector, $u\in\R^m$ the input vector, $y\in\R^p$ the output vector, $w\in\R^d$ the disturbance vector and the output measurement is corrupted by a noise $\eta(k)\in\R^p$. Also, $A_x\in\R^{n\times n}$, $B_u\in\R^{n\times m}$, $D_w\in\R^{n\times d}$, $B_\sigma\in\R^{n\times \nu}$, $\tilde A\in\R^{\nu\times n}$, $\tilde B\in\R^{\nu\times m}$, 
$\tilde D\in\R^{\nu\times d}$, $C\in\R^{p\times n}$, and $\sigma(\cdot)=[\,\sigma_1(\cdot)\;\dots\;\sigma_\nu(\cdot)\,]^{\top}$ is a decentralized vector of scalar functions satisfying the following assumption.
\begin{ass}\label{ass:sigmoid_function}
    Each component $\sigma_i:\R\rightarrow\R$, $i=1,\dots,\nu$, is a sigmoid function, i.e. a bounded, twice continuously differentiable function with positive first derivative at each point and one and only one inflection point in $v_i=0$.
    Also, $\sigma_i(\cdot)$ is Lipschitz continuous with unitary Lipschitz constant and such that $\sigma_i(0)=0$, $\frac{  \partial\sigma_i(v_i)}{\partial v_i}\big|_{v_i=0}=1$ and $\sigma_i(v_i)\in[-1,1]$, $\forall v_i\in\R$.\hfill{}$\square$
\end{ass}
We make the following assumptions on the boundedness of the terms $w(k)$ and $\eta(k)$.
\begin{ass}\label{ass:disturbance}
The disturbance $w(k)$ satisfies, for all $k\geq 0$, $w(k)\in \mathcal{E}(Q_w^0)$, where $Q_w^0\in\mathbb S_+^d$.\hfill{}$\square$
\end{ass}
\begin{ass}\label{ass:output_noise}
    The disturbance $\eta(k)$ satisfies, for all $k\geq 0$, $\eta(k)\in \mathcal{E}(Q_\eta^0)$, where $Q_\eta^0\in\mathbb S_+^p$.\hfill{}$\square$
\end{ass}
%
The goal of this work is to design a control system that allows us to steer the plant's output to a neighbour of a constant setpoint $\bar{y}\in\R^p$ that must respect the following assumption. 
\begin{ass}\label{ass:output_reference}
The output reference $\bar{y}$ must be selected in such a way that there exist $\bar x$ and $\bar u$ such that $\bar x=A_x\bar x+B_u\bar u+B_\sigma \sigma(\tilde A\bar x+\tilde B\bar u)$ and $\bar y=C\bar x$.\hfill{}$\square$
\end{ass}
\subsection{Model reformulation and incremental sector condition}
As done in~\cite{la2025regional}, for convenience the plant dynamics \eqref{eq:plant_dynamics} is rewritten as follows
\begin{equation}\label{eq:plant_dynamics_q_open_loop}
    \begin{cases}
        x(k+1)=Ax(k)+Bu(k)+Dw(k)+B_qq(v(k))\\
        v(k)=\tilde Ax(k)+\tilde Bu(k)+\tilde Dw(k)\\
        y(k)=Cx(k)+\eta(k).
    \end{cases}
\end{equation}
where $A=A_x+B_\sigma \tilde A$, $B=B_u+B_\sigma \tilde B$, $D=D_w+B_\sigma \tilde D$, $B_q=-B_\sigma$, and the vector $ q(v)=\begin{bmatrix}q_1(v_1) & \dots &q_\nu(v_\nu)\end{bmatrix}^\top\in\R^\nu$ is composed of the functions $q_i(v_i)=v_i-\sigma(v_i)$, $i=1,\dots,\nu$.\\ 
%
A key property used in this work is the incremental sector condition satisfied by $q_i(\cdot)$, proved in the following lemma.
\begin{lemma}\label{lem:sector_condition}
Under Assumption~\ref{ass:sigmoid_function}, for all $h_i>1$, $\exists\bar v_i(h_i)\in\R_{+}$ such that, for all pairs $(v_i,v_i+\Delta v_i)\in[-\bar v_i(h_i),\bar v_i(h_i)]^2$, it holds that
\begin{equation}\label{eq:sector_condition_h}
\Delta q_i(\Delta v_i-h_i\Delta q_i)\geq 0,
\end{equation}
where $\Delta q_i=q_i(v_i+\Delta v_i)-q_i(v_i)$. Function $\bar v_i(h_i):(1,+\infty)\to(0,+\infty)$ is a continuous, strictly decreasing function such that $\bar v_i(h_i) \to +\infty$ as $h_i\to 1^+$ and $\bar v_i(h_i)\to0$ as $h_i\to +\infty$. Also, in case $h_i=1$, condition \eqref{eq:sector_condition_h} holds for all $(v_i,v_i+\Delta v_i)\in\R^2$.\hfill{}$\square$
\end{lemma}
The proof of Lemma~\ref{lem:sector_condition} is provided in Appendix~\ref{appendix_A}.\footnote{
The role of~\eqref{eq:sector_condition_h} is that of characterising the nonlinearity appearing in model~\eqref{eq:plant_dynamics_q_open_loop}. This characterisation, through a suitable matrix reformulation, will be used to provide a design condition, which will apply to all models where the nonlinearity $q_i(\cdot)$, for each $i=1,\dots,\nu$ enjoys~\eqref{eq:sector_condition_h}. For instance, if we set $h_i=1$,~\eqref{eq:sector_condition_h} implies that, for all $v_i$ and $\Delta v_i$, $\Delta q_i=q_i(v_i+\Delta v_i)-q_i(v_i)$ and $\Delta v_i-\Delta q_i=\sigma_i(v_i+\Delta v_i)-\sigma_i(v_i)$ have the same sign.  
However, (global) stability properties may not be enforced for all models, requiring to provide a local characterisation of the nonlinearity. 
Indeed, as proved in Lemma~\ref{lem:sector_condition}, in view of the properties of $\sigma_i(\cdot)$, $\Delta q_i/\Delta v_i \leq 1/h_i$ (from which~\eqref{eq:sector_condition_h} essentially follows) for any possible $h_i> 1$ in case $v_i$ and $v_i+\Delta v_i$ both lie in the interval $[-\bar v_i(h_i),\bar v_i(h_i)]$ and where $\bar v_i(h_i)$ is a function of $h_i$.\\ 
Parameters $h_i$, for $i=1,\dots,\nu$ will take the role of further degrees of freedom of our controller design problem. 
Generally speaking, by increasing $h_i$ we progressively relax the structural conditions required for the feasibility of our design problem but, at the same time, we restrict the region of validity of the stability properties. 
Therefore, for each possible set of values of $h_i$, $i=1,\dots,\nu$, 
to be determined as part of the solution to the controller design problem, we can directly compute the corresponding validity region.}\smallskip\\
Note that the analytic expression of $\bar v_i(h_i)$ cannot be defined explicitly. However, for any given $h_i>1$, we can compute $\bar v_i(h_i)$ numerically by solving \eqref{opt:locality_constraints} (see Appendix~\ref{appendix_A}).\\
For notational reasons, for each possible $H=\diag(h_1, \dots,h_\nu)$ we can define the set
\[\mathcal{I}(H)=\{i\in\{1,\dots,\nu\}\ : \ h_i>1\}.\]
which is basically the set of indexes of sector conditions for which a local characterization is necessary to make the control design problem feasible. 
Letting $\Delta v=[\,\Delta v_1,\;\dots,\;\Delta v_\nu\,]^\top$ and $\Delta q(v+\Delta v,v)=q(v+\Delta v)-q(v)$,~\eqref{eq:sector_condition_h} can be reformulated, in matrix form as
\begin{equation}\label{eq:sector_inequality}
    \Delta q(v+\Delta v,v)^\top S(\Delta v-H\Delta q(v+\Delta v,v))\geq 0,  \forall v,v+\Delta v\in\mathcal{V}(H)
\end{equation}
which is valid for any matrix $S\in\mathbb{D}_+^\nu$ and where the set $\mathcal{V}(H)$ is defined as
\begin{equation*}
    \mathcal{V}(H)\coloneqq\{v\in\R^\nu \ : \ v_i\in[-\bar v_i(h_i),\bar v_i(h_i)], \ \forall i\in \mathcal{I}(H)\}.
\end{equation*}
%
%
\section{Nonlinear observer design}\label{sec:observer}
In many real-world applications, particularly when working with RNN models, where state variables may not directly correspond to measurable physical quantities, the system state is often unavailable. 
For this reason, to estimate the system state $x(k)$, which is required for the proper definition of the output-feedback scheme, a nonlinear observer is proposed of the form
\begin{equation}\label{eq:observer_dynamics}
    \begin{cases}
        \hat{x}(k+1) = A\hat{x}(k) + Bu(k) + L(y(k) - C\hat{x}(k)) + B_q q(\hat{v}(k)) \\
        \hat{v}(k) = \tilde{A} \hat{x}(k) + \tilde{B} u(k) + \tilde{L} (y(k) - C\hat{x}(k))
    \end{cases}
\end{equation}
where $\hat x\in\R^n$ is the observer state and $L\in\R^{n\times p}$ and $\tilde L\in\R^{\nu\times p}$ are observer gains.
The following proposition provides a design condition for the observer \eqref{eq:observer_dynamics}.
\begin{proposition}
    \label{prop:dISS_observer}
    Consider the observer dynamics \eqref{eq:observer_dynamics}, define the estimation error $e(k)\coloneq x(k)-\hat x(k)$, and let Assumptions \ref{ass:sigmoid_function}, \ref{ass:disturbance}, and \ref{ass:output_noise} hold. If there exist $\gamma_\mathrm{o}\in\R_+$, and matrices $P_\mathrm{o},Q_{\mathrm{o},x}\in \mathbb S^n_+$, $Q_{o,w_\mathrm{o}}\in \mathbb S^{d+p}_+$, $H_\mathrm{o},S_\mathrm{o}=\diag(s_{\mathrm o,1},\dots,s_{\mathrm o,\nu})\in\mathbb{D}_+^v$, with $H_\mathrm{o}\succeq I_\nu$, $N\in\R^{n\times p}$, and $\tilde N\in\R^{\nu\times p}$, such that 
    
   \begin{subequations}\label{eq:LMIs_observer}    \begin{equation}\label{eq:dISS_condition_obs}
    \begin{bmatrix}
        P_\mathrm{o}-Q_{\mathrm{o},x}&-\!\tilde A^\top S_\mathrm{o}+C^\top\tilde N^\top&0&A^\top P_\mathrm{o}-C^\top N^\top\\
        -S_\mathrm{o}\tilde A+\tilde NC&2S_\mathrm{o}H_\mathrm{o}&\begin{bmatrix}-S_\mathrm{o}\tilde D&\tilde N\end{bmatrix}&B_q^\top P_\mathrm{o}\\
        0&\begin{bmatrix}-\tilde D^\top S_\mathrm{o}\\\tilde N^\top\end{bmatrix}&Q_{o,w_\mathrm{o}}&\begin{bmatrix}D^\top P_\mathrm{o}\\N^\top\end{bmatrix}\\
        P_\mathrm{o}A-NC&P_\mathrm{o}B_q&\begin{bmatrix}P_\mathrm{o}D&N\end{bmatrix}&P_\mathrm{o}
    \end{bmatrix}\!\!\succeq\!0,
    \end{equation}
    \begin{equation}\label{eq:mRPI_set_condition_1_dISS_obs}
        Q_{o,w_\mathrm{o}}\preceq \diag{(Q_w^0/2,Q_\eta^0/2)},
    \end{equation}
    \begin{equation}\label{eq:mRPI_set_condition_2_dISS_obs}
        Q_{\mathrm{o},x}-P_\mathrm{o}/\gamma_\mathrm{o}\succeq 0,
    \end{equation}     
    \begin{equation}\label{eq:locality_set_dISS_obs1}
    \begin{bmatrix}
        P_\mathrm{o}/(2\gamma_\mathrm{o})&0&\tilde A_{i}^\top\\
        0&Q_{w}^0/2&\tilde D_{i}^\top\\
        \tilde A_{i}&\tilde D_{i}&\bar v_i(h_{\mathrm{o},i})^2
    \end{bmatrix}\succeq 0, \ \forall i\in\mathcal{I}(H_\mathrm{o}),
    \end{equation}
    \begin{equation}\label{eq:locality_set_dISS_obs2}
    \begin{bmatrix}
        P_\mathrm{o}/(2\gamma_\mathrm{o})&0&C^\top\tilde N_i^\top\\
        0&Q_\eta^0/2&\tilde N_i^\top\\
        \tilde N_iC&\tilde N_{i}&\bar v_i(h_{\mathrm{o},i})^2s_{\mathrm{o},i}^2
        \end{bmatrix}\succeq 0, \ \forall i\in\mathcal{I}(H_\mathrm{o}),
    \end{equation}
\end{subequations}
then, setting $L=P_\mathrm{o}^{-1}N$ and $\tilde L=S_\mathrm{o}^{-1}\tilde N$,
\begin{itemize}
        \item if $\mathcal{I}(H_\mathrm{o})=\emptyset$,  \eqref{eq:observer_dynamics} has ISS observer error dynamics with respect to the sets $\R^n$, $\mathcal{E}(Q_w^0)$, and $\mathcal{E}(Q_\eta^0)$;
        \item if $\mathcal{I}(H_\mathrm{o})\neq\emptyset$ and if the pair ($\hat x$,$u$) satisfies \begin{equation}\label{eq:locality_condition_obs}
        \tilde A \hat x+\tilde Bu\in\bar{\mathcal V}_{\mathrm{o}},
        \end{equation}
        where the set
        \begin{equation*}
        \bar{\mathcal V}_{\mathrm{o}}=\left(\mathcal V(H_\mathrm{o})\ominus\left(\tilde A\mathcal E(P_\mathrm{o}/\gamma_\mathrm{o})\oplus\tilde D\mathcal{E}(Q_w^0)\right)\right)\cap\left(\mathcal V(H_\mathrm{o})\ominus\left(\tilde LC\mathcal E(P_\mathrm{o}/\gamma_\mathrm{o})\oplus\tilde L\mathcal{E}(Q_\eta^0)\right)\right)
        \end{equation*}
        is non-empty,
        then \eqref{eq:observer_dynamics} has ISS observer error dynamics with respect to the sets $\mathcal{E}(P_\mathrm{o}/\gamma_\mathrm{o})$, $\mathcal{E}(Q_w^0)$, and $\mathcal{E}(Q_\eta^0)$.
    \end{itemize} 
Moreover, the set $\mathcal{E}(P_\mathrm{o}/\gamma_\mathrm{o})$ is an RPI set for the observer error dynamics, i.e., if $e(0)\in\mathcal{E}(P_\mathrm{o}/\gamma_\mathrm{o})$, then $e(k)\in\mathcal{E}(P_\mathrm{o}/\gamma_\mathrm{o})$ for all $k\geq 0$.\hfill{}$\square$
\end{proposition}
The proof of Proposition~\ref{prop:dISS_observer} is provided in Appendix~\ref{appendix_A}.\smallskip\\
Regarding the existence of a feasible solution to problem \eqref{eq:LMIs_observer}, note that the feasibility of conditions \eqref{eq:mRPI_set_condition_1_dISS_obs}-\eqref{eq:locality_set_dISS_obs2} primarily depends on the magnitude of the disturbances $w(k)$ and $\eta(k)$. On the other hand, the feasibility of \eqref{eq:dISS_condition_obs} is more critical, as it is directly linked to the structure of model~\eqref{eq:plant_dynamics_q_open_loop}, as stated in the following lemma.
\begin{lemma}\label{lem:NScond_obs}
  A necessary and sufficient condition for the existence of $H_\mathrm{o}\in\mathbb D_+^\nu$, with $H_\mathrm{o}\succeq I_\nu$, ensuring feasibility of \eqref{eq:dISS_condition_obs} is that the pair $(A,C)$ is detectable.\hfill{}$\square$
\end{lemma}
The proof of Lemma~\ref{lem:NScond_obs} is provided in Appendix~\ref{appendix_A}.\smallskip\\
%
\section{Controller design}\label{sec:local_dISS}
\subsection{The control law}
In this section, we define the control law
\begin{equation}\label{eq:control_law_tracking} 
u(k)=\bar u+K(\hat{x}(k)-\bar x),
\end{equation}
where  $K\in\R^{m\times n}$ is the control gain, which is here designed in order to confer stability properties to the control system.\\
First note that the dynamics of the estimated state $\hat x(k)$ satisfies the following 
\begin{equation}\label{eq:observer_dynamics_2}
\begin{cases}
\hat x(k+1)=A\hat x(k)+Bu(k)+LCe(k)+L\eta(k)+B_qq(\hat v(k))\\
\hat v(k)=\tilde{A}\hat x(k)+\tilde{B}u(k)+\tilde LCe(k)+\tilde L \eta(k)
\end{cases}
\end{equation}
Let us define the exogenous variable vector $w_\mathrm{c}(k)=[\eta(k)^\top,\, e(k)^\top\,]^\top$, acting as a disturbance in \eqref{eq:observer_dynamics_2}. 
Introducing matrices $A_\mathrm{K}=A+BK$, $\tilde A_\mathrm{K}=\tilde A+\tilde BK$, $D_{\mathrm{c}}=[\,L\; LC\,]$, and $\tilde D_{\mathrm{c}}=[\,\tilde L\; \tilde LC\,]$, the closed-loop observer state evolves according to
\begin{equation}\label{eq:closed_loop_observer}
\begin{cases}
\hat x(k+1)=A_\mathrm{K} \hat x(k)+B(\bar u-K\bar x)+D_{\mathrm{c}} w_\mathrm{c}(k)+B_q q(\hat v(k))\\
\hat v(k)=\tilde A_\mathrm{K}\hat x(k)+\tilde B(\bar u-K\bar x)+\tilde D_{\mathrm{c}} w_\mathrm{c}(k)\\
\end{cases}
\end{equation}
The following theorem establishes a sufficient condition for the $\delta$ISS of the observer state dynamics \eqref{eq:closed_loop_observer}.
\\
\begin{proposition}\label{prop:dISS}
Let Assumptions \ref{ass:sigmoid_function}, \ref{ass:disturbance}, and \ref{ass:output_noise} hold. Also, assume that $e(k)\in\mathcal{E}(P_\mathrm{o}/\gamma_\mathrm{o})$ for $k\geq0$, and define $Q_{w_\mathrm{c}}^0=\diag{(Q_\eta^0,P_\mathrm{o}/\gamma_\mathrm{o})}$. If there exist $\gamma_\mathrm{c}\in\R_{+}$, and matrices $Q_{\mathrm{c}},\tilde Q_{\mathrm{c},x}\in\mathbb{S}_+^n$, $Q_{c,w_\mathrm{c}}\in\mathbb{S}_+^d$ and $H_\mathrm{c},U_\mathrm{c}\in\mathbb D_+^\nu$, with $H_\mathrm{c}\succeq I_\nu$, and $Z\in\R^{m,n}$ that satisfy the following conditions
\begin{subequations}\label{eq:LMIs}
\begin{equation}\label{eq:dISS_condition}
\begin{bmatrix}
    Q_{\mathrm{c}}-\tilde Q_{\mathrm{c},x}&-Q_{\mathrm{c}}\tilde A^\top\!-\!Z^\top \tilde B^\top &0&Q_\mathrm{c}A^\top+Z^\top B^\top\\
    -\tilde AQ_\mathrm{c}\!-\!\tilde B Z&2H_\mathrm{c}U_\mathrm{c}& -\tilde D_{\mathrm{c}}&U_\mathrm{c}B_q^\top\\
    0&-\tilde D_{\mathrm{c}}^\top &Q_{c,w_\mathrm{c}}&D_{\mathrm{c}}^\top\\
    AQ_\mathrm{c}+BZ&B_qU_\mathrm{c}&D_{\mathrm{c}}&Q_{\mathrm{c}}
\end{bmatrix}
    \succeq 0,
\end{equation}
\begin{equation}\label{eq:mRPI_set_condition_1_dISS}
Q_{c,w_\mathrm{c}}\preceq Q_{w_\mathrm{c}}^0,
\end{equation}
\begin{equation}\label{eq:mRPI_set_condition_2_dISS}
 \tilde Q_{\mathrm{c},x}-Q_{\mathrm{c}}/\gamma_\mathrm{c}\succeq0,
\end{equation}
\begin{equation}\label{eq:locality_set_dISS}
\begin{bmatrix}\
        Q_{\mathrm{c}}/(2\gamma_\mathrm{c})&0&Q_{\mathrm{c}}\tilde A_{i}^\top+Z^\top\tilde B_i^\top\\
        0&Q_{w_\mathrm{c}}^0/2&\tilde D_{\mathrm{c},i}^\top\\
        \tilde A_{i}Q_{\mathrm{c}}+\tilde B_iZ&\tilde D_{\mathrm{c},i}&(\bar v_i(h_{\mathrm{c},i})-|\tilde A_i\bar x+\tilde B_i\bar u|)^2
        \end{bmatrix}\succeq 0, \quad \forall i\in\mathcal{I}(H_\mathrm{c}),
\end{equation}
\begin{equation}\label{eq:locality_set_dISS_ctrobs1}
\begin{aligned}
&\begin{bmatrix}\
        Q_{\mathrm{c}}/(3\gamma_\mathrm{c})&0&0&Q_{\mathrm{c}}\tilde A_{i}^\top+Z^\top\tilde B_i^\top\\
        0&P_\mathrm{o}/(3\gamma_\mathrm{o})&0&\tilde A_{i}^\top\\
        0&0&Q_w^0/3&\tilde D_{i}\\
        \tilde A_{i}Q_{\mathrm{c}}+\tilde B_iZ&\tilde A_{i}&\tilde D_{i}&(\bar v_i(h_{\mathrm{o},i})-|\tilde A_i\bar x+\tilde B_i\bar u|)^2
        \end{bmatrix}\succeq 0, \quad \forall i\in\mathcal{I}(H_\mathrm{o}),
\end{aligned}
\end{equation}
\begin{equation}\label{eq:locality_set_dISS_ctrobs2}
\begin{aligned}
&\begin{bmatrix}\
        Q_{\mathrm{c}}/(3\gamma_\mathrm{c})&0&0&Q_{\mathrm{c}}\tilde A_{i}^\top+Z^\top\tilde B_i^\top\\
        0&P_\mathrm{o}/(3\gamma_\mathrm{o})&0& C^\top \tilde L_{i}^\top\\
        0&0&Q_\eta^0/3&\tilde L_{i}^\top\\
        \tilde A_{i}Q_{\mathrm{c}}+\tilde B_iZ&\tilde L_{i}C&\tilde L_{i}&(\bar v_i(h_{\mathrm{o},i})-|\tilde A_i\bar x\!+\!\tilde B_i\bar u|)^2
        \end{bmatrix}\succeq 0, \quad \forall i \in \mathcal{I}(H_\mathrm{o}),
\end{aligned}
\end{equation}
\begin{equation}\label{eq:locality_eq_dISS}
\bar v_i(h_{\mathrm{c},i})\geq|\tilde A_i\bar x+\tilde B_i\bar u|, \quad \forall i\in\mathcal{I}(H_\mathrm{c}),
\end{equation}
\begin{equation}\label{eq:locality_eq_dISS_ctrobs}
\bar v_i(h_{\mathrm{o},i})\geq|\tilde A_i\bar x+\tilde B_i\bar u|, \quad \forall i\in\mathcal{I}(H_\mathrm{o}),
\end{equation}
\end{subequations}
then, setting $K=ZQ_\mathrm{c}^{-1}$ and $P_\mathrm{c}=Q_{\mathrm{c}}^{-1}$, 
\begin{itemize}
    \item if $\mathcal{I}(H_\mathrm{c})=\emptyset$, the system described by \eqref{eq:closed_loop_observer} is $\delta$ISS with respect to the sets $\R^n$ and $\mathcal E(Q_{w_\mathrm{c}}^0)$;
    \item if $\mathcal{I}(H_\mathrm{c})\neq\emptyset$, system \eqref{eq:closed_loop_observer} is $\delta$ISS with respect to the sets $\mathcal{E}(P_\mathrm{c}/\gamma_\mathrm{c})\oplus\bar x$ and $\mathcal E(Q_{w_\mathrm{c}}^0)$. 
\end{itemize} 
Moreover, the set $\mathcal{E}(P_\mathrm{c}/\gamma_\mathrm{c})\oplus\bar x$ is an RPI set for the closed-loop observer dynamics and, if $\hat x(0)\in\mathcal{E}(P_\mathrm{c}/\gamma_\mathrm{c})\oplus\bar x$, then also \eqref{eq:locality_condition_obs} holds for all $k>0$.\hfill{}$\square$
\end{proposition}
The proof of Proposition~\ref{prop:dISS} is provided in Appendix~\ref{appendix_A}.\smallskip\\
It is important to consider also the feasibility of the LMIs \eqref{eq:LMIs}. Similarly to the case of \eqref{eq:LMIs_observer}, while the feasibility of \eqref{eq:mRPI_set_condition_1_dISS}-\eqref{eq:locality_eq_dISS_ctrobs} basically depends on the magnitude of the output noise $\eta(k)$ and of the estimation error $e(k)$ and on the choice of the equilibrium setpoint, inequality
\eqref{eq:dISS_condition} is the key one since it is essentially related to the system structure. In the following lemma we provide a sufficient and necessary condition for the existence of a suitable $H_\mathrm{c}$ that allows us to provide a feasible solution to it.
\begin{lemma}\label{lem:NScond}
  A necessary and sufficient condition for the existence of $H_\mathrm{c}\in\mathbb D_+^\nu$, with $H_\mathrm{c}\succeq I_\nu$, ensuring feasibility of \eqref{eq:dISS_condition} is that $(A,B)$ is stabilizable.\hfill{}$\square$
\end{lemma}
The proof of Lemma~\ref{lem:NScond} is provided in Appendix~\ref{appendix_A}.
\subsection{The control system}\label{subsec:control_system}
\begin{figure}[tp]
    \fontsize{8}{12}\selectfont
    \centering
    \def\svgwidth{0.5\columnwidth}
\begingroup%
  \makeatletter%
  \providecommand\color[2][]{%
    \errmessage{(Inkscape) Color is used for the text in Inkscape, but the package 'color.sty' is not loaded}%
    \renewcommand\color[2][]{}%
  }%
  \providecommand\transparent[1]{%
    \errmessage{(Inkscape) Transparency is used (non-zero) for the text in Inkscape, but the package 'transparent.sty' is not loaded}%
    \renewcommand\transparent[1]{}%
  }%
  \providecommand\rotatebox[2]{#2}%
  \newcommand*\fsize{\dimexpr\f@size pt\relax}%
  \newcommand*\lineheight[1]{\fontsize{\fsize}{#1\fsize}\selectfont}%
  \ifx\svgwidth\undefined%
    \setlength{\unitlength}{894.99993896bp}%
    \ifx\svgscale\undefined%
      \relax%
    \else%
      \setlength{\unitlength}{\unitlength * \real{\svgscale}}%
    \fi%
  \else%
    \setlength{\unitlength}{\svgwidth}%
  \fi%
  \global\let\svgwidth\undefined%
  \global\let\svgscale\undefined%
  \makeatother%
  \begin{picture}(1,0.50614529)%
    \lineheight{1}%
    \setlength\tabcolsep{0pt}%
    \put(0,0){\includegraphics[width=\unitlength]{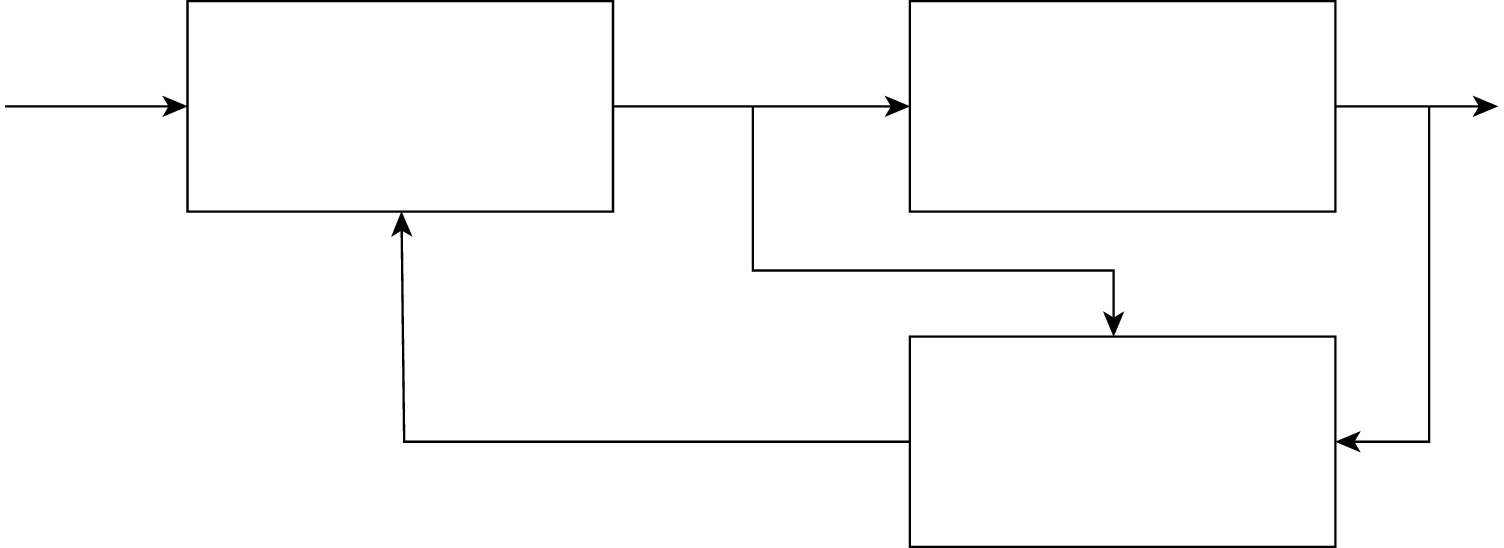}}%
    \put(0.16,0.285){\color[rgb]{0,0,0}\makebox(0,0)[lt]{\lineheight{1.25}\smash{\begin{tabular}[t]{l}$\bar u+K(\hat x-\bar x)$\end{tabular}}}}%
    \put(0.5,0.32){\color[rgb]{0,0,0}\makebox(0,0)[lt]{\lineheight{1.25}\smash{\begin{tabular}[t]{l}$u$\end{tabular}}}}%
    \put(0.5,0.085){\color[rgb]{0,0,0}\makebox(0,0)[lt]{\lineheight{1.25}\smash{\begin{tabular}[t]{l}$\hat x$\end{tabular}}}}%
    \put(0.94,0.33){\color[rgb]{0,0,0}\makebox(0,0)[lt]{\lineheight{1.25}\smash{\begin{tabular}[t]{l}$y$\end{tabular}}}}%
    \put(0.01,0.32){\color[rgb]{0,0,0}\makebox(0,0)[lt]{\lineheight{1.25}\smash{\begin{tabular}[t]{l}$(\bar{x},\bar{u})$\end{tabular}}}}%
    \put(0.7,0.285){\color[rgb]{0,0,0}\makebox(0,0)[lt]{\lineheight{1.25}\smash{\begin{tabular}[t]{l}Plant\end{tabular}}}}%
    \put(0.67,0.068){\color[rgb]{0,0,0}\makebox(0,0)[lt]{\lineheight{1.25}\smash{\begin{tabular}[t]{l}Observer\end{tabular}}}}%
  \end{picture}%
\endgroup%

    \caption{Output-feedback control architecture with static control law.}
    \label{fig:control_architecture}
\end{figure}

We now consider the control system depicted in Figure~\ref{fig:control_architecture}, which integrates
the observer dynamics~\eqref{eq:observer_dynamics} with the control law~\eqref{eq:control_law_tracking}. To provide stability properties, the observer gains $(L, \tilde{L})$ and the control gain $K$ must be designed according to Propositions~\ref{prop:dISS_observer} and~\ref{prop:dISS}, respectively.\\
Note, however, that conditions \eqref{eq:LMIs_observer} and \eqref{eq:LMIs} are not LMIs and the solution could result numerically complex.  
To avoid this, we propose the following LMI-based heuristic procedure.\smallskip\\
\emph{Step 1}: Ensure that the pair $(A,C)$ is observable and the pair $(A,B)$ is controllable, and initialise $H_\mathrm o=H_\mathrm c=I_{\nu}$ and $\gamma_\mathrm o=\gamma_\mathrm c=1$.\smallskip\\
\emph{Step 2}: Solve the restricted LMI problem \eqref{eq:dISS_condition_obs}-\eqref{eq:locality_set_dISS_obs1} with decision variables $P_\mathrm{o}$, $Q_{\mathrm{o},x}$, $Q_{\mathrm o,w}$, $N$, $\tilde{N}$, and $S_\mathrm{o}$.\smallskip\\
\emph{Step 3}: If a solution is not found, update
\[ H_\mathrm o, \gamma_\mathrm o =\mathrm{parameters\_update}\,(H_\mathrm o, \gamma_\mathrm o, \gamma_\mathrm {max},\epsilon_h,\epsilon_\gamma),\]
 where $\gamma_\mathrm {max}$, $\epsilon_h$ and $\epsilon_\lambda$ are design parameters, and return to Step 2.\smallskip\\
\emph{Step 4}: Fix $S_\mathrm{o}$ and solve \eqref{eq:LMIs_observer} as an LMI problem. If feasible, set $L = P_\mathrm{o}^{-1}N$ and $\tilde{L} = S_\mathrm{o}^{-1}\tilde{N}$, otherwise, return to Step 3.\smallskip\\
\emph{Step 5}: Solve the LMI problem \eqref{eq:dISS_condition}-\eqref{eq:locality_set_dISS_ctrobs2} with decision variables $P_\mathrm{c}$, $Q_{\mathrm{c},x}$, $Q_{\mathrm c,w}$, $Z$, and $S_\mathrm{c}$.\smallskip\\
\emph{Step 6}: If a solution is not found, update 
\[H_\mathrm c, \gamma_\mathrm c =\mathrm{parameters\_update}\,(H_\mathrm c, \gamma_\mathrm c, \gamma_\mathrm {max},\epsilon_h,\epsilon_\gamma),\]
 otherwise set $K = ZQ_\mathrm{c}^{-1}$, $P_\mathrm{c} = Q_{\mathrm{c}}^{-1}$.\smallskip\\
Steps~$2$-$4$ compute the observer gains and the associated RPI set $\mathcal{E}(P_\mathrm{o}/\gamma_\mathrm{o})$ based on Proposition~\ref{prop:dISS_observer}. Since \eqref{eq:LMIs_observer} is not an LMI problem, we make an initial guess for $H_\mathrm{o}$ and $\gamma_\mathrm{o}$ and solve the restricted LMI problem \eqref{eq:dISS_condition_obs}-\eqref{eq:locality_set_dISS_obs1}. This provides an initial partial solution. Then, by fixing $S_\mathrm{o}$, we solve \eqref{eq:LMIs_observer} as an LMI problem to compute the final solution.  If the problem is infeasible, $H_\mathrm{o}$ and $\gamma_\mathrm{o}$ are updated iteratively until a feasible solution is found.\\ 
Once the observer gains are determined, a similar procedure is applied in Steps~$5$–$6$ to determine the control gain $K$ and the associated RPI set $\mathcal{E}(P_\mathrm{c}/\gamma_\mathrm{c}) \oplus \bar{x}$.\\
To ensure that the solution exists in the largest possible region, in Step~1 we initialise $H_\mathrm{c} = I_\nu$ and $H_\mathrm{o} = I_\nu$ and attempt to solve \eqref{eq:LMIs} (resp., \eqref{eq:LMIs_observer}) by incrementally increasing $\gamma_\mathrm{c}$ (resp., $\gamma_\mathrm{o}$). If no feasible solution is found, we search for a local solution by updating $H_\mathrm{c}$ and/or $H_\mathrm{o}$ according to Algorithm~\ref{alg:design_procedure}. Note, however, that if the disturbances are too large or if the region where stability is enforced is too small (i.e. if \eqref{eq:locality_eq_dISS} or \eqref{eq:locality_eq_dISS_ctrobs} are not satisfied), no solution may exist for the selected setpoint.\\
Also, note that the proposed procedure allows us to formulate the design problem as a sequence of convex problems, but it is, in general, not optimal. Less conservative solutions may be obtained by increasing $H_\mathrm{o}$ and $H_\mathrm{c}$ non-uniformly. In this case, one can also leverage model simulations on the dataset to identify which directions are more suitable for reducing the region in which stability is enforced. \\
Alternatively to Algorithm~\ref{alg:design_procedure}, a joint design of $L$, $\tilde{L}$, and $K$ by simultaneously solving \eqref{eq:LMIs_observer} and \eqref{eq:dISS_condition}–\eqref{eq:locality_set_dISS_ctrobs2} would likely yield a less conservative solution. However, this approach leads to a bilinear matrix inequality problem, which is more computationally intensive and prone to numerical issues.\\
Finally, note that, as we aim to maximise the region of attraction $\mathcal{E}(P_\mathrm{c}/\gamma_\mathrm{c})$ of the controller (resp., minimise the estimation error set $\mathcal{E}(P_\mathrm{o}/\gamma_\mathrm{o})$), we recommend complementing the LMI problem \eqref{eq:dISS_condition}-\eqref{eq:locality_set_dISS_ctrobs2} in Step $5$ (resp., \eqref{eq:LMIs_observer} in Step $4$) with a cost function $f_c(Q_{\mathrm{c}})$ (resp., $f_\mathrm{o}(P_\mathrm{o})$). These functions should be selected by the designer to appropriately shape the RPI sets (see~\cite{boyd1994linear} for a detailed discussion of possible functions).
\begin{algorithm}[t]
\caption{parameters\_update}
\label{alg:design_procedure}
\begin{algorithmic}[1]
\State \textbf{Require} $H\in\mathbb D_+^\nu$, and $\gamma,\gamma_\text{max},\varepsilon_h, \varepsilon_\gamma \in \mathbb{R}_+$
\State \textbf{Ensure} $H$, $\gamma$
    \If{$\gamma < \gamma_\text{max}$}
        \State \textbf{Set} $\gamma = \gamma + \varepsilon_\gamma$
    \Else
        \State \textbf{Set} $H = H + \varepsilon_h I_\nu$, $\gamma = 1$
    \EndIf
\end{algorithmic}
\end{algorithm}
The properties of the resulting closed-loop system are stated next.
\begin{theorem}\label{th:closed_loop_K_as_gain}
   Suppose that $w(k)$ and $\eta(k)$ satisfy Assumptions~\ref{ass:disturbance} and \ref{ass:output_noise}, respectively. Then, if the design procedure described in Steps~$1$-$6$ admits a feasible solution, and if $e(0) \in \mathcal{E}(P_o/\gamma_o)$ and $\hat{x}(0) \in \mathcal{E}(P_c/\gamma_c)\oplus \bar x$, there exist class $\mathcal{K}$-functions $\gamma_w$ and $\gamma_\eta$ such that    \begin{equation}\label{eq:closed_loop_K_as_gain}
        \overline{\lim}_{k \to+ \infty} \norm{x(k) - \bar{x}} \leq \gamma_w( \overline{\lim}_{k \to+ \infty} \norm{w(k)} )
        + \gamma_\eta (\overline{\lim}_{k \to+ \infty} \norm{\eta(k)}).
    \end{equation}
    \hfill{}$\square$
\end{theorem}
The proof of Theorem~\ref{th:closed_loop_K_as_gain} is provided in Appendix~\ref{appendix_A}.\smallskip\\
Theorem~\ref{th:closed_loop_K_as_gain} guarantees that if the system state is correctly initialised, the output $y$ is robustly driven toward a neighbourhood of the desired setpoint $\bar{y}$, whose size depends only on the magnitude of the disturbances affecting the system.
%
%
%
%
%
\section{MPC control design}\label{sec:MPC}
\begin{figure}[tp]
    \fontsize{8}{12}\selectfont
    \centering
    \def\svgwidth{0.47\columnwidth}
    \input{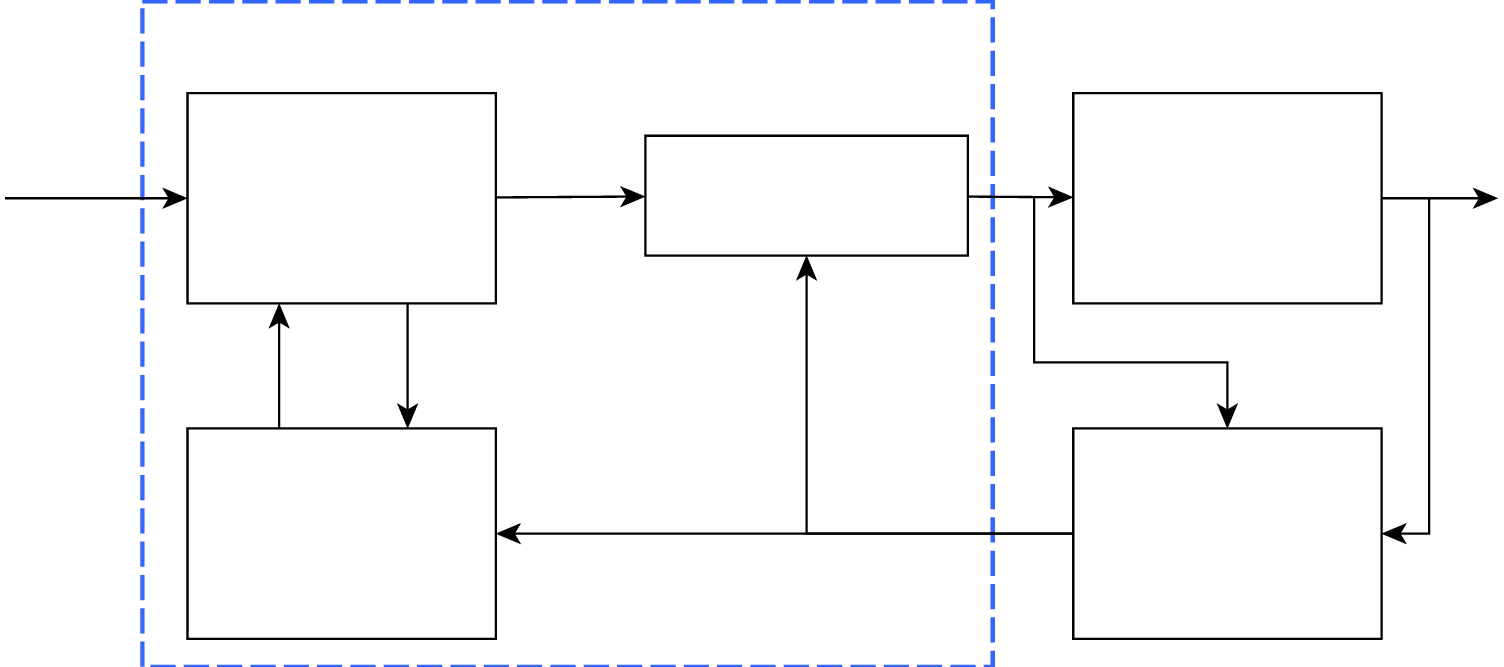}
    \caption{NMPC output-feedback control architecture.}
    \label{fig:control_architecture_MPC}
\end{figure}

Proposition~\ref{prop:dISS}, discussed in Section \ref{sec:local_dISS}, provides a condition for the design of a static output-feedback control law \eqref{eq:control_law_tracking} for tracking the setpoint triple $(\bar x,\bar u,\bar y)$.
A potential weakness of the control system described in the previous section is related to the local $\delta$ISS property: indeed, convergence is ensured when the initial conditions lie in suitable, but possibly small, RPI sets, which may limit the applicability of the controller around the selected nominal operating conditions. To overcome this potential problem, in this section we design a suitable MPC controller that uses \eqref{eq:control_law_tracking} as an auxiliary control law; in this case the invariant set computed in the previous section just serves as a terminal feasible set, potentially greatly enlarging the region of attraction of the setpoint.\smallskip\\
As customary in the MPC framework, in this section we assume that the system operation must be subject to input and output constraints, i.e. $(u,y)\in\mathbb U\times\mathbb Y$ and we consider the following.
\begin{ass}\label{ass:compact_sets}
    The sets $\mathbb U$ and $\mathbb Y$ are polytopes, i.e., $\mathbb{U}=\{u\in\R^m:G_{\mathrm{u}}u\leq b_{\mathrm{u}}\}$, where $G_{\mathrm{u}}\in\R^{n_s\times m}$ and $b_{\mathrm{u}}\in\R^{n_s}$, and $\mathbb{Y}=\{y\in\R^p:G_{\mathrm{y}}y\leq b_{\mathrm{y}}\}$, where $G_{\mathrm{y}}\in\R^{n_r\times p}$ and $b_{\mathrm{y}}\in\R^{n_r}$. \hfill{}$\square$
\end{ass}
To address the stated control objectives, we propose the NMPC output-feedback control architecture represented in Figure \ref{fig:control_architecture_MPC}, which consists of the nonlinear observer~\eqref{eq:observer_dynamics}, designed based on Proposition~\ref{prop:dISS_observer} and a tube-based state-feedback NMPC law.
\subsection{The nominal prediction model}
The model used for computing the state predictions in the MPC-related optimization problem is obtained from \eqref{eq:observer_dynamics_2} by neglecting the uncertain terms depending on $\eta(k)$ and $e(k)$. The resulting nominal system dynamics is
\begin{equation}\label{eq:nom_system}
    \tilde x(k+1) = A\tilde x(k)+B\tilde u(k)+B_qq(\tilde A\tilde x(k)+\tilde B\tilde u(k)),
\end{equation}
where $\tilde x\in\R^n$ and $\tilde u\in\R^m$ represent the nominal state and nominal input vectors, respectively.\\
The input $\tilde u(k)$ and the state $\tilde x(k)$ will be used to compute the real plant input through the following control law
\begin{equation}\label{eq:control_law_tube}
u(k)=\tilde u(k)+K\tilde e(k),
\end{equation}
where $\tilde e(k)=\hat x(k)-\tilde x(k)$, and the gain $K$ is selected according to the following proposition.
%
%
%
%
%
%
Note that, applying \eqref{eq:control_law_tube} to the observer dynamics \eqref{eq:observer_dynamics_2}, the resulting closed-loop dynamics evolves as
\begin{equation}\label{eq:closed_loop_observer_MPC}
\hat x(k+1)=A_\mathrm{K}\hat x(k)+B(\tilde u(k)-K\tilde x(k))+D_{\mathrm{c}}w_\mathrm{c}(k)+B_qq(\tilde{A}_K\hat x(k)+\tilde{B}(\tilde u(k)-K\tilde x(k))+\tilde D_{\mathrm{c}}w_\mathrm{c}(k)).
\end{equation}
\begin{proposition}\label{prop:dISS_control}
    Let Assumptions \ref{ass:sigmoid_function}, \ref{ass:disturbance}, and \ref{ass:output_noise} hold. Also, assume that $e(k)\in\mathcal{E}(P_\mathrm{o}/\gamma_\mathrm{o})$, for $k\geq0$. If there exist $\gamma_\mathrm{c}\in\R_+$, and matrices $Q_{\mathrm{c}},\tilde Q_{\mathrm{c},x}\in \mathbb S^n_+$, $Q_{c,w_\mathrm{c}}\in \mathbb S^{n+p}_+$, $H_\mathrm{c},U_\mathrm{c}\in\mathbb{D}_+^v$, with $H_\mathrm{c}\succeq I_\nu$, and $Z\in\R^{m\times n}$ that satisfy \eqref{eq:dISS_condition}-\eqref{eq:mRPI_set_condition_2_dISS}, and such that  \begin{equation}\label{eq:locality_set_dISS_mpc}
    \begin{bmatrix}\
        Q_{\mathrm{c}}/(2\gamma_\mathrm{c})&0&Q_{\mathrm{c}}\tilde A_{i}^\top+Z^\top \tilde B_i^\top\\
        0\!\!&\!\!Q_{w_\mathrm{c}}^0/2&\tilde D_{\mathrm{c},i}\\
        \tilde A_{i}Q_{\mathrm{c}}+\tilde B_iZ&\tilde D_{\mathrm{c},i}&\bar v_i(h_{\mathrm{c},i})^2
        \end{bmatrix}\succeq 0, \ \forall i\in\mathcal{I}(H_\mathrm{c}),
    \end{equation}
then, setting $K=ZQ_\mathrm{c}^{-1}$ and $P_\mathrm{c}=Q_{\mathrm{c}}^{-1}$, \begin{itemize}
        \item if $\mathcal{I}(H_\mathrm{c})=\emptyset$, the closed-loop observer dynamics \eqref{eq:closed_loop_observer_MPC} is $\delta$ISS with respect to the sets $\R^n$ and $\mathcal{E}(Q_{w_\mathrm{c}}^0)$;
        \item if $\mathcal{I}(H_\mathrm{c})\neq\emptyset$ and the pair ($\tilde x$, \ $\tilde u$) satisfies 
        \begin{equation}\label{eq:locality_constr_MPC}
        \tilde A \tilde x+\tilde B\tilde u\in\bar{\mathcal V}_{\mathrm{c}},
        \end{equation}
        where the set
        \begin{equation*}
        \bar{\mathcal V}_{\mathrm{c}}=\mathcal V(H_\mathrm{c})\ominus\left(\tilde A_\mathrm{K}\mathcal E(P_\mathrm{c}/\gamma_\mathrm{c})\oplus\tilde D_{\mathrm{c}}\mathcal{E}(Q_{w_\mathrm{c}}^0)\right)
        \end{equation*}
        is non-empty, the closed-loop observer dynamics \eqref{eq:closed_loop_observer_MPC} is $\delta$ISS with respect to the sets $\mathcal{E}(P_\mathrm{c}/\gamma_\mathrm{c})\oplus\tilde x$ and $\mathcal{E}(Q_{w_\mathrm{c}}^0)$.
    \end{itemize} 
Moreover, the set $\mathcal{E}(P_\mathrm{c}/\gamma_\mathrm{c})$ is an RPI set for the dynamics of $\tilde e(k)$.\hfill{}$\square$
\end{proposition}
The proof of Proposition~\ref{prop:dISS_control} is provided in Appendix~\ref{appendix_A}.\smallskip\\
Note that the state $x(k)$ of the real system is computed as $x(k)=\tilde x(k)+e(k)+\tilde e(k)$. Therefore, if the observer gains $(L,\tilde L)$ and the control gain $K$ are selected according to Propositions~\ref{prop:dISS_observer} and~\ref{prop:dISS_control}, respectively, the set $\mathcal{E}(P_\mathrm{o}/\gamma_\mathrm{o})\oplus\mathcal{E}(P_\mathrm{c}/\gamma_\mathrm{c})$ is an RPI set for the dynamics of $x(k)-\tilde x(k)$, provided that $e(0)\in\mathcal{E}(P_\mathrm{o}/\gamma_\mathrm{o})$ and, if $\mathcal{I}(H_\mathrm{o})\neq\emptyset$ or $\mathcal{I}(H_\mathrm{c})\neq\emptyset$, condition \eqref{eq:locality_condition_obs} and \eqref{eq:locality_constr_MPC}, respectively, hold. The following lemma reformulates \eqref{eq:locality_condition_obs} with respect to the state and input of the nominal system.
\begin{lemma}\label{lem:MPC_locality_cstr} 
Let the assumptions of Proposition~\ref{prop:dISS_control} hold. If there exist $\gamma_\mathrm{c}\in\R_+$, $Q_{\mathrm{c}}\in\mathbb S_+^n$ and $Z\in\R^{m\times n}$ that, for all $i\in\mathcal I(H_\mathrm{o})$, satisfy 
\begin{subequations}
\begin{equation}\label{eq:locality_set_dISS_obs_mpc_1}
    \begin{bmatrix}
    P_\mathrm{o}/(3\gamma_\mathrm{o})&0&0&\tilde A_i^\top\\
    0&Q_{\mathrm{c}}/(3\gamma_\mathrm{c})&0&Q_{\mathrm{c}}\tilde A_{i}^\top+Z\tilde B_i^\top\\
    0&0&Q_w^0/3&\tilde D_i^\top\\
    \tilde A_i&\tilde A_{i}Q_{\mathrm{c}}+\tilde B_iZ&\tilde D_i&\bar v(h_{\mathrm{o},i})^2
    \end{bmatrix}\!\!\succeq 0
    \end{equation}
    \begin{equation}\label{eq:locality_set_dISS_obs_mpc_2}
    \begin{bmatrix}
    P_\mathrm{o}/(3\gamma_\mathrm{o})&0&0&C^\top\tilde L_i^\top\\
    0&Q_{\mathrm{c}}/(3\gamma_\mathrm{c})&0&Q_{\mathrm{c}}\tilde A_{i}^\top+Z\tilde B_i^\top\\
    0&0&Q_\eta^0/3&\tilde L_i^\top\\
    \tilde L_iC&\tilde A_{i}Q_{\mathrm{c}}+\tilde B_iZ&\tilde L_i&\bar v(h_{\mathrm{o},i})^2
    \end{bmatrix}\succeq 0
    \end{equation}
\end{subequations}
\normalsize
then, setting $K=ZQ_\mathrm{c}^{-1}$ $P_\mathrm{c}=Q_{\mathrm{c}}^{-1}$, if $\tilde e\in\mathcal{E}(P_\mathrm{c}/\gamma_\mathrm{c})$ and the pair $(\tilde x,\tilde u)$ satisfies
    \begin{equation}\label{eq:locality_constr_obsMPC}
    \tilde A\tilde x+\tilde B\tilde u\in\mathcal{\bar V}_{\mathrm{o}_\mathrm{c}},
\end{equation}
where the set 
\begin{equation*}
\mathcal{\bar V}_{\mathrm{o}_\mathrm{c}}=\left(\mathcal V(H_\mathrm{o})\ominus\left(\tilde A\mathcal E(P_\mathrm{o}/\gamma_\mathrm{o})\oplus\tilde D\mathcal{E}(Q_w^0)\oplus\tilde A_\mathrm{K}\mathcal E(P_\mathrm{c}/\gamma_\mathrm{c})\right)\right)\cap
\left(\mathcal V(H_\mathrm{o})\ominus\left(\tilde LC\mathcal E(P_\mathrm{o}/\gamma_\mathrm{o})\oplus\tilde L\mathcal{E}(Q_\eta^0)\oplus\tilde A_\mathrm{K}\mathcal E(P_\mathrm{c}/\gamma_\mathrm{c})\right)\right),
\end{equation*}
is non-empty, then \eqref{eq:locality_condition_obs} holds. \hfill{}$\square$
\end{lemma}
The proof of Lemma~\ref{lem:MPC_locality_cstr} is provided in Appendix~\ref{appendix_A}.
\subsection{Constraint tightening}
The formulation of the NMPC problem requires expressing the state and control constraints in terms of equivalent constraints for the nominal system.
We make the following assumption.
\begin{ass}\label{ass:set}
        There exist $\epsilon_u,\epsilon_y\in\R_+$, $\bar{\bar y}\in\R^p$ and $\bar{\bar u}\in\R^m$
    such that $C\mathcal E(P_\mathrm{c}/\gamma_\mathrm{c})\oplus C\mathcal E(P_\mathrm{o}/\gamma_\mathrm{o})\oplus\mathcal{B}_{\epsilon_y}^{(p)}(\bar{\bar y})\subseteq \mathbb{Y}$ and $K\mathcal E(P_\mathrm{c}/\gamma_\mathrm{c})\oplus\mathcal{B}_{\epsilon_u}^{(m)}(\bar{\bar u})  \subseteq \mathbb{U}$.\hfill{}$\square$
\end{ass}
In view of Assumption~\ref{ass:set}, 
we can define the tightened sets $\tilde{\mathbb Y}=\mathbb Y\ominus C\left(\mathcal{E}(P_\mathrm{o}/\gamma_\mathrm{o})\oplus\mathcal{E}(P_\mathrm{c}/\gamma_\mathrm{c})\right)$ and $\tilde{\mathbb U}=\mathbb U\ominus K\mathcal{E}(P_\mathrm{c}/\gamma_\mathrm{c})$.
Moreover, to guarantee \eqref{eq:locality_constr_MPC}  when $\mathcal{I}(H_\mathrm{c})\neq\emptyset$, we introduce the set 
\[\mathcal{V}_{\mathrm{c}}(H_\mathrm{c})=\{v\in\R^\nu:|v_i|\leq\bar v_{\mathrm{c},i}(h_{\mathrm{c},i}), \ \forall i\in \mathcal{I}(H_\mathrm{c})\}\]
where
\begin{equation*}
\bar{v}_{\mathrm{c},i}(h_{\mathrm{c},i}) = \bar{v}_i(h_{\mathrm{c},i}) - \max_{\substack{\tilde e \in \mathcal{E}(P_\mathrm{c}/\gamma_\mathrm{c}), \\ w_\mathrm{c} \in \mathcal{E}(Q_{w_\mathrm{c}})}} |\tilde{A}_{K,i} \tilde e  + \tilde{D}_{\mathrm{c},i} w_\mathrm{c}|.
\end{equation*}
Note that $\mathcal{V}_{\mathrm{c}}(H_\mathrm{c})\subseteq\bar{\mathcal V}_{\mathrm{c}}$.
Additionally, to guarantee \eqref{eq:locality_constr_obsMPC}, we define the set
\[
\mathcal{V}_\mathrm{o}(H_\mathrm{o})=\{v\in\R^\nu:|v_i|\leq\bar v_{\mathrm{o},i}^\star(h_{\mathrm{o},i}), \\\ \forall i\in \mathcal{I}(H_\mathrm{o})\}
\]
where $\bar v_{\mathrm{o},i}^\star(h_{\mathrm{o},i})=\min(\bar v_{\mathrm{o},i}(h_{\mathrm{o},i}),\tilde v_{\mathrm{o},i}(h_{\mathrm{o},i}))$, with
\begin{equation*}
\bar{v}_{\mathrm{o},i}(h_{\mathrm{o},i}) = \bar{v}_i(h_{\mathrm{o},i}) -\max_{\substack{e \in \mathcal{E}(P_\mathrm{o}/\gamma_\mathrm{o}), \\ w \in \mathcal{E}(Q_w^0), \\ \tilde e \in \mathcal{E}(P_\mathrm{c}/\gamma_\mathrm{c})}}  |\tilde{A}_{i} e + \tilde{D}_{i} w+\tilde A_{\mathrm{K},i}  \tilde e|,
\end{equation*}
and
\begin{equation*}
\tilde{v}_{\mathrm{o},i}(h_{\mathrm{o},i}) = \bar{v}_i(h_{\mathrm{o},i}) - \max_{\substack{e \in \mathcal{E}(P_\mathrm{o}/\gamma_\mathrm{o}), \\ \eta \in \mathcal{E}(Q_\eta^0), \\ \tilde e \in \mathcal{E}(P_\mathrm{c}/\gamma_\mathrm{c})}} |\tilde{L}_{i}C e + \tilde{L}_{i} \eta+\tilde A_{\mathrm{K},i} \tilde e|.
\end{equation*}
Note that $\mathcal{V}_\mathrm{o}(H_\mathrm{o})\subseteq\bar{\mathcal V}_{o_c}$.\\
Finally, we modify Assumption~\ref{ass:output_reference} as follows.
\begin{ass}\label{ass:output_reference_constr}
The output reference $\bar{y}\in \mathcal{INT}(\tilde{\mathbb Y})$ must be selected in such a way that the equilibrium triple $(\bar x,\bar u,\bar y)$ satisfies $\bar u\in \mathcal{INT}(\tilde{\mathbb{U}})$ and $\tilde A\bar x+\tilde B\bar u\in\mathcal{V}_\mathrm{o}(H_\mathrm{o})\cap\mathcal{V}_{\mathrm{c}}(H_\mathrm{c})$.
\end{ass}
\subsection{Tube-based NMPC design}
In line with the classic tube-based NMPC, at each time instant $k$ a finite-horizon optimal control problem (FHOCP) is stated with reference to the nominal system \eqref{eq:nom_system}. The control input sequence $\tilde{u}([k:k+N-1])$ on a given prediction horizon of length $N$ and the nominal state $\tilde{x}(k)$ are considered as optimization variables.\\
The FHOCP is formulated as follows
\begin{subequations}\label{opt:NMPC}
\begin{align}
&\min_{\tilde{x}(k),\tilde{u}([k:k+N-1])} J(\tilde{x}([k:k+N]),\tilde{u}([k:k+N-1]) \notag\\
&\text{subject to:} \notag\\
&\hat x(k)-\tilde x(k) \in \mathcal{E}(P_\mathrm{c}/\gamma_\mathrm{c})\label{opt:NMPC_init}\\
&\forall\tau=0,\dots,\ N-1:\notag\\
&\quad \begin{aligned}
&\tilde{x}(k+\tau+1) = A \tilde{x}(k+\tau) + B \tilde{u}(k+\tau) \\
&\quad + B_q q \left( \tilde{A} \tilde{x}(k+\tau) + \tilde{B} \tilde{u}(k+\tau) \right)
\end{aligned} \label{opt:NMPC_system} \\
&\quad \tilde{u}(k+\tau)\in \tilde{ \mathbb U} \label{opt:NMPC_input}\\
&\quad C\tilde{x}(k+\tau)\in\tilde{\mathbb Y}\label{opt:NMPC_outputcstr}\\
&\quad\tilde A\tilde x(k+\tau)+ \tilde B\tilde u(k+\tau)\in\mathcal{V}_{\mathrm{c}}(H_\mathrm{c})\label{opt:NMPC_localcstr_control}\\
&\quad\tilde A\tilde x(k+\tau)+ \tilde B\tilde u(k+\tau)\in\mathcal{V}_\mathrm{o}(H_\mathrm{o})\label{opt:NMPC_localcstr_observer}\\
&\tilde{x}(k+N) \in \mathbb X_\mathrm{f}(\bar x) \label{opt:NMPC_terminal}
\end{align}
\end{subequations}
Note that constraint \eqref{opt:NMPC_system} embeds the dynamics of the predictive model, which is initialized by constraint \eqref{opt:NMPC_init} in the neighbourhood of the state estimate $\hat x(k)$. Input and output constraints are imposed through \eqref{opt:NMPC_input} and \eqref{opt:NMPC_outputcstr}, respectively. Locality constraints \eqref{eq:locality_constr_MPC} and \eqref{eq:locality_constr_obsMPC} are enforced through \eqref{opt:NMPC_localcstr_control} and \eqref{opt:NMPC_localcstr_observer}, respectively. Finally, the nominal state is required to reach the terminal set $\mathbb X_\mathrm{f}(\bar x)$ at the end of the prediction horizon, as specified by \eqref{opt:NMPC_terminal}.  Let the cost function, which penalizes the deviation of $(\tilde x,\ \tilde u)$ from the target equilibrium $(\bar x,\ \bar u)$, be defined by 
\begin{equation*}
    J=\sum_{\tau=0}^{N-1}\left(\norm{\tilde{x}(k+\tau)-\bar x}_{\Lambda_x}^2+\norm{\tilde{u}(k+\tau)-\bar u}_{\Lambda_u}^2\right)+V_\mathrm{f}(\tilde{x}(k+N),\bar x)
\end{equation*}
where $\Lambda_x\in\mathbb S_+^n$, $\Lambda_u\in\mathbb S_+^m$, and $V_\mathrm{f}$ is the terminal cost.\\
According to the receding horizon approach, the optimal solution of the FHOCP, denoted as $(\tilde x(k),\ \tilde u([k:k+N]))$, is computed at each time step, and the control law~\eqref{eq:control_law_tube} is applied to the system.
\subsection{Terminal ingredients}
As customary, we base the design of the NMPC terminal components, namely the terminal set $\mathbb{X}_\mathrm{f}(\bar x)$ and the terminal cost $V_\mathrm{f}(\tilde x(k),\bar x)$, on the definition of an auxiliary law. In particular, we select as an auxiliary law 
\begin{equation}\label{eq:control_law_aux}
    \tilde u(k)=\bar u+K(\tilde x(k)-\bar x),
\end{equation}
The terminal set is defined as 
\[\mathbb X_\mathrm{f}(\bar x)=\mathcal E(P_\mathrm{f}/\gamma_\mathrm{f})\oplus\bar x,\]
where $P_\mathrm{f}\in\mathbb S_+^n$ and $\gamma_\mathrm{f}\in\R_+$. The terminal cost is defined as \[V_\mathrm{f}(\tilde x(k+N),\bar x)=\norm{\tilde x(k+N)-\bar x}^2_{P_\mathrm{f}}.\]
%
\subsection{Design procedure}\label{sec:NMPC_design}
Consider the control system represented in Figure~\ref{fig:control_architecture_MPC}. The following procedure is proposed for the design of the observer gains $(L,\tilde L)$, the control gain $K$, and the terminal ingredients $\mathbb{X}_\mathrm{f}(\bar x)$ and $V_\mathrm{f}(\tilde x(K+N),\bar x)$.\smallskip\\
\step{1f} To determine the observer gains $(L,\tilde L)$  and the control gain $K$, execute the design procedure presented in Section~\ref{subsec:control_system}, replacing Step~$5$ with the LMI problem
\begin{align*}
&\min_{\substack{Q_{\mathrm{c}},\tilde Q_{\mathrm{c},x}\in\mathbb{S}_+^n, Q_{c,w_\mathrm{c}}\in\mathbb{S}_+^{n+p},\\Z\in\R^{m\times n},U_\mathrm{c}\in\mathbb{D}_+^\nu}} \beta_{\mathrm{c}}\\
&\text{subject to:}\\
&\quad\eqref{eq:dISS_condition}–\eqref{eq:mRPI_set_condition_2_dISS},\eqref{eq:locality_set_dISS_mpc},\eqref{eq:locality_set_dISS_obs_mpc_1},\eqref{eq:locality_set_dISS_obs_mpc_2},\\
&\quad f_{LMI,\mathrm{c}}(Q_{\mathrm{c}})\preceq\beta_{\mathrm{c}}I
\end{align*}
where $f_{LMI,\mathrm{c}}(Q_{\mathrm{c}})$ is a function defined by the designer to minimize the RPI set $\mathcal{E}(P_\mathrm{c}/\gamma_\mathrm{c})$ (see \cite{boyd1994linear}).\\
\step{2f} Define the tightened sets $\tilde{\mathbb Y}$ and $\tilde{\mathbb U}$, the sets $\mathcal{V}_{\mathrm{c}}(H_\mathrm{c})$ and $\mathcal{V}_\mathrm{o}(H_\mathrm{o})$, and set $H_\mathrm{f}=I_\nu$.\\  
\step{3f} Solve the following optimization problem
\begin{subequations}\label{opt:terminal_ingredients}
\begin{align}
&\min_{\tilde{\gamma}_{\mathrm{f}}\in\R_+,P_\mathrm{f}\in\mathbb S_+^n,S_{\mathrm{f}}\in\mathbb D_+^\nu} \beta_{{\mathrm{f}},1}\text{trace}(P_\mathrm{f})+\beta_{{\mathrm{f}},2}\gamma_\mathrm{f}\label{eq:cost_terminal}\\
&\text{subject to:}\notag\\
&
\begin{bmatrix}
P_\mathrm{f}-\Lambda_x-K^\top\Lambda_uK & -\tilde A_{\mathrm{K}}^\top S_{\mathrm{f}} & A_{K}^\top P_\mathrm{f}\\
-S_{\mathrm{f}}\tilde A_{K} & 2S_{\mathrm{f}}H_{\mathrm{f}} & B_q^\top P_\mathrm{f}\\
P_{\mathrm{f}}A_{K} & P_{\mathrm{f}}B_q & P_\mathrm{f}
\end{bmatrix}\succeq 0\label{eq:dISS_terminal}\\
&
\begin{bmatrix}
    P_\mathrm{f}&C^\top G_{\mathrm{y},r}^\top\\
    G_{\mathrm{y},r}C&\tilde\gamma_\mathrm{f}(b_{\mathrm{y},r}-G_{\mathrm{y},r}C\bar x)^2
\end{bmatrix}\succeq0, \ \forall r=1,\dots,\ n_r\label{eq:output_constr_terminal}\\
&
\begin{bmatrix}
    P_\mathrm{f}&K^\top G_{\mathrm{u},s}^\top\\
    G_{\mathrm{u},s}K&\tilde\gamma_\mathrm{f}(b_{\mathrm{u},s}-G_{\mathrm{u},s}\bar u)^2
\end{bmatrix}\succeq0, \ \forall s=1,\dots,\ n_s\label{eq:input_constr_terminal}\\
&
\begin{bmatrix}
    P_\mathrm{f}&\tilde A_{\mathrm{K},i}^\top\\
    \tilde A_{\mathrm{K},i}&\tilde\gamma_\mathrm{f}v_i(h_{\mathrm{f},i})^2
\end{bmatrix}\succeq0, \ \forall i\in \mathcal I(H_\mathrm{f})\label{eq:locality_constr_hf_terminal}\\
&
\begin{bmatrix}
    P_\mathrm{f}&\tilde A_{\mathrm{K},i}^\top\\
    \tilde A_{\mathrm{K},i}&\tilde\gamma_\mathrm{f}(\bar v_{\mathrm{c},i}(h_{\mathrm{c},i})-|\tilde A_i\bar{x}+\tilde B_i\bar{u}|)^2
\end{bmatrix}\succeq0, \ \forall i\in \mathcal I(H_\mathrm{c})\label{eq:locality_constr_hc_terminal}\\
&
\begin{bmatrix}
    P_\mathrm{f}&\tilde A_{\mathrm{K},i}^\top\\
    \tilde A_{\mathrm{K},i}&\tilde\gamma_\mathrm{f}(\bar v_{\mathrm{o},i}^\star(h_{\mathrm{o},i})-|\tilde A_i\bar{x}+\tilde B_i\bar{u}|)^2
\end{bmatrix}\succeq0, \ \forall i\in \mathcal I(H_\mathrm{o})\label{eq:locality_constr_ho_terminal}
\end{align}
\end{subequations}
where $\beta_{{\mathrm{f}},1},\beta_{{\mathrm{f}},2}\in\R_+$ are design parameters. If \eqref{opt:terminal_ingredients} is feasible, set $\gamma_\mathrm{f}=1/\tilde{\gamma}_{\mathrm{f}}$, otherwise set $H_\mathrm{f}=H_\mathrm{f}+\epsilon_hI_\nu$ and repeat Step~3f.\\
Note that, for the NMPC design, we aim to obtain the tightest outer approximation of the minimal RPI set, as specified in Step 1f, in order to reduce the conservativeness introduced by constraint tightening. Conversely, we seek to maximise the NMPC terminal set $\mathbb X_\mathrm{f}(\bar x)$ through the minimisation of the cost function \eqref{eq:cost_terminal} in \eqref{opt:terminal_ingredients}.
\subsection{Main result}
The theoretical properties of the resulting closed-loop system are summarized in the following theorem.
\begin{theorem}\label{th:FHOCP_guarantees}
Suppose that Assumptions \ref{ass:sigmoid_function}-\ref{ass:output_reference_constr} are verified and that $e(0) \in \mathcal{E}(P_\mathrm{o}/\gamma_\mathrm{o})$. Then, if the FHOCP  \eqref{opt:NMPC} admits a solution at time $k=0$, it admits a solution for each $k>0$. Moreover, the output ${y}$ asymptotically converges to $\bar y\oplus C\mathcal{E}(P_\mathrm{c}/\gamma_\mathrm{c})\oplus C\mathcal{E}(P_\mathrm{o}/\gamma_\mathrm{o})$ as $k\rightarrow+\infty$, \textcolor{blue}and the constraints $(u(k),y(k))\in\mathbb U\times\mathbb Y$ are fulfilled for all $k\geq0$.\hfill{}$\square$
\end{theorem}
The proof of Theorem~\ref{th:FHOCP_guarantees} is provided in Appendix~\ref{appendix_A}.\smallskip\\
This result ensures that, for any feasible initial state, the proposed method steers the nominal system state to the desired equilibrium $\bar x$, while satisfying the tightened constraints. Furthermore, since the state observer \eqref{eq:observer_dynamics} and the control law \eqref{eq:control_law_tube} guarantee that $x(k)-\tilde x(k)$ remains bounded for all $k\in\mathbb  Z_{\geq0}$, the output $y$ of the perturbed system is robustly driven towards a neighbourhood of $\bar y$ while satisfying input and output constraints.
%
%
\section{Numerical Results}\label{sec:simulations}
\begin{figure}[htbp]
     \centering
    \includegraphics[width=0.7\columnwidth]{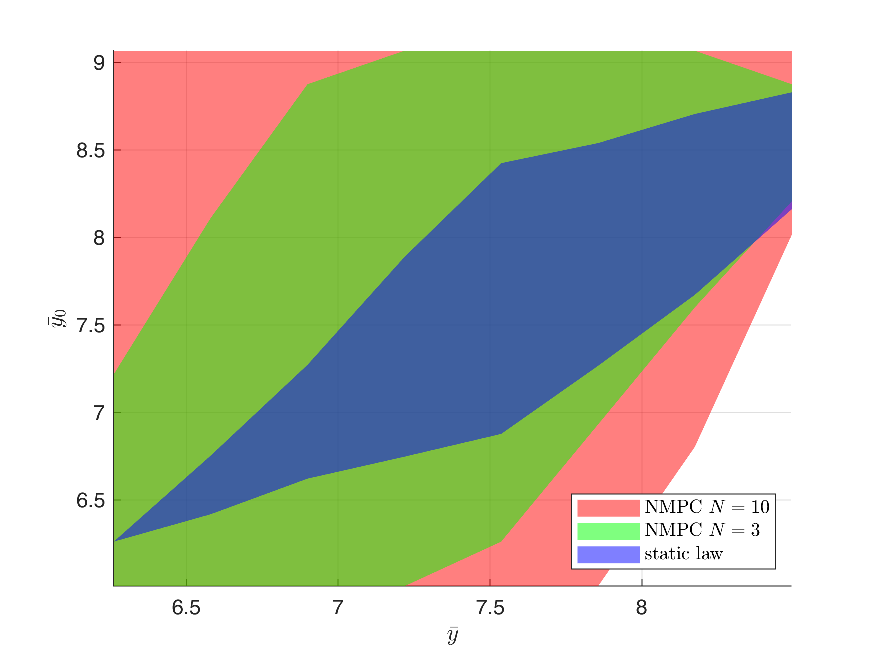}
    \caption{Range of initial conditions expressed in terms of the output $y$ from which $\bar{y}$ is reachable. The figure compares the static control law and the NMPC law with $N=3$ and $N=10$.}
      \label{fig:output_range}
\end{figure}

\begin{figure}[htbp]
     \centering
    \includegraphics[width=0.7\columnwidth]{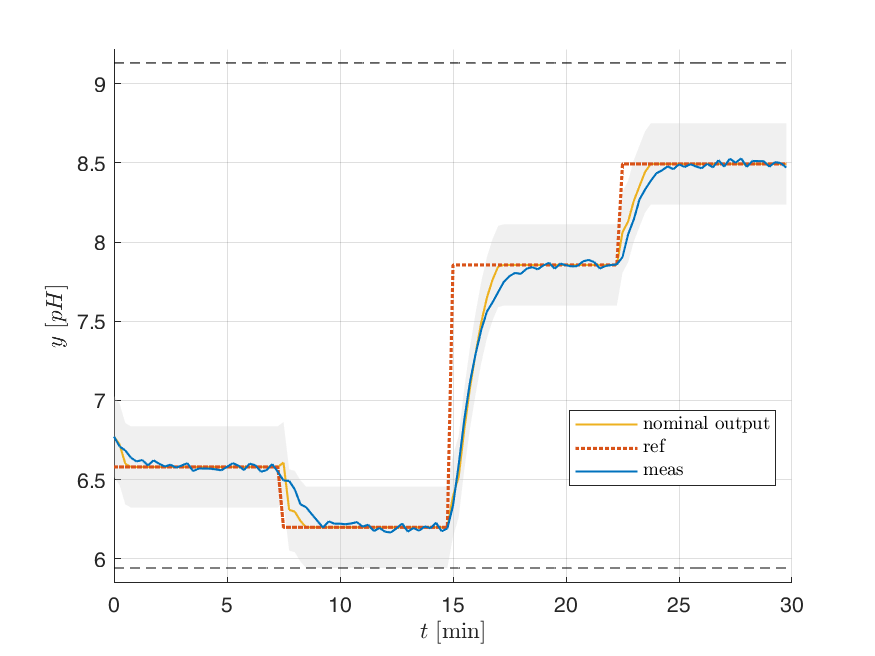}
    \caption{Closed-loop output performance. 
    Black dashed lines denote output constraints, while shaded regions represent tubes around the nominal trajectories.}
      \label{fig:output_cl}
\end{figure}

\begin{figure}[htbp]
     \centering
    \includegraphics[width=0.7\columnwidth]{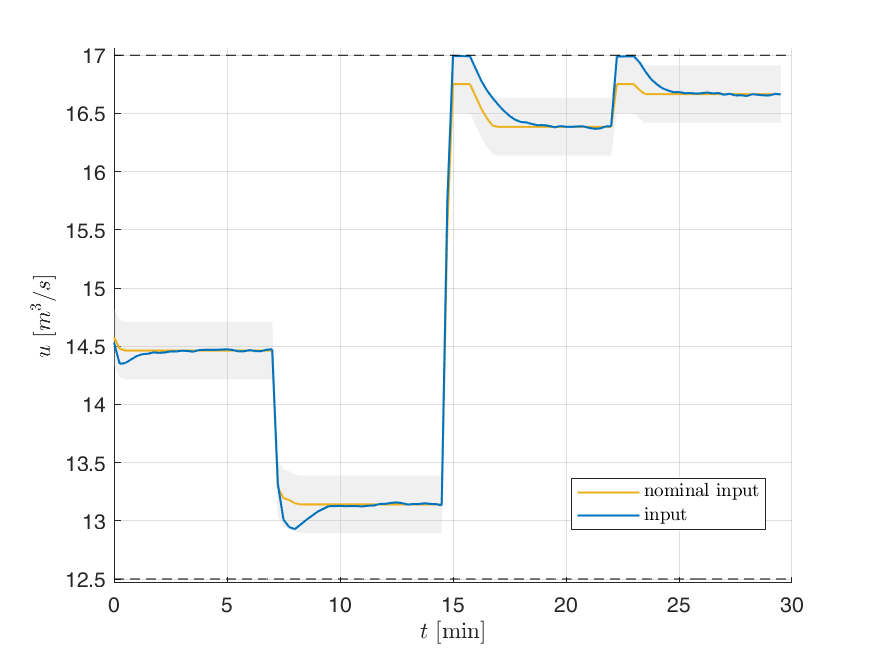}
    \caption{Evolution of the control input. 
    Black dashed lines denote input constraints, while shaded regions represent tubes around the nominal trajectories.}
      \label{fig:input_cl}
\end{figure}

In this section, the theoretical results are validated by applying the proposed approach to control a simulation model of a pH-neutralisation process \cite{hall1989modelling}.\\ 
The overall model is a nonlinear SISO system, where the controllable input $u$ is the inlet alkaline
base flow rate, and the measured output $y$ is
the pH of the output flow rate. Input and output are subject to saturation limits, i.e. $u=[11.2,\ 17.2]$ and $y=[5.94,\ 9.13]$.\\
A first-principles model of the pH-neutralisation process, as described in \cite{hall1989modelling}, has been implemented in MATLAB. An input-output data sequence was collected with a sampling time of $15$~s by feeding the simulator with a multilevel pseudo-random signal, designed to excite the system across different operating regions. 
This dataset has been subsequentely normalised and then used to identify the parameters of an RNN-based nonlinear model of the class~\eqref{eq:nom_system}, characterised by $n=8$ states, and with $\sigma_i=\tanh(\cdot)$ for $i=1,\dots,5$.\smallskip\\
Based on the identified model, the two output-feedback schemes, one using the static control law presented in Section~\ref{sec:local_dISS} and the other based on NMPC proposed in Section \ref{sec:MPC}, have been implemented in the MATLAB environment and compared for controlling the simulated pH-neutralisation process. To align with the framework of this paper we consider an output noise $\eta(k)\in[-0.01, 0.01]$.\\ The static control law-based scheme was designed using the procedure presented in Section~\ref{subsec:control_system}. Specifically, to ensure constraint satisfaction under the static control law, we designed the control gain with an associated RPI set contained within the constraints, i.e. $C(\mathcal{E}(P_\mathrm{c}/\gamma_\mathrm{c})\oplus\mathcal{E}(P_\mathrm{o}/\gamma_\mathrm{o}))\oplus\bar y\in\mathbb Y$ and $K(\mathcal{E}(P_\mathrm{c}/\gamma_\mathrm{c})\oplus\bar u)\in\mathbb U$. This was achieved by adding suitable LMIs into Step~$5$, following a procedure similar to that in the proof of Theorem \ref{th:FHOCP_guarantees}. For the NMPC-based scheme, we followed the design procedure outlined in Section~\ref{sec:NMPC_design}, using the same gain originally designed for the static law to ensure a fair comparison between the two schemes. The FHOCP was solved using CasADi \cite{andersson2019casadi}, with an average computational time of $0.34$ seconds per control step. 
Figure~\ref{fig:output_range} illustrates, for each setpoint $\bar{y}$, the set of initial setpoints $\bar{y}_0$ from which the control system can track $\bar y$ without violating constraints, assuming that the system has reached the desired steady state when the setpoint change occurs. Specifically, the figure compares the static control law, the NMPC law with a prediction horizon of $N=3$, and the NMPC law with $N=10$. Note that, we compute the region of attraction of the controllers with respect to the output rather than the states because the states of the RNN are generally not associated with relevant physical quantities. On the other hand, setpoint variations are often of greater interest from a practical point of view.  The results show that the region of attraction of the static control law is smaller than the region of attraction of NMPC. Therefore, the static control law-based scheme may not be sufficient in applications where large setpoint variations may occur. Moreover, as shown in the figure, the NMPC law significantly enlarges the set dimension as $N$ increases. \\
Finally, the closed-loop performance of the output-feedback NMPC control architecture, with $N=10$, has been tested in tracking a piecewise constant reference signal where each setpoint variation is contained in the red set in Figure~\ref{fig:output_range}.
Figures~\ref{fig:output_cl}-\ref{fig:input_cl} display the closed-loop simulation results. The figures show that the controller achieves satisfactory tracking accuracy while fulfilling the
constraints on $u$ and $y$. Also, note that, despite the presence of the output disturbance and the state estimation error, the closed-loop input and output trajectories lie within the tubes $\tilde u(k)\oplus K\mathcal{E}(P_\mathrm{c}/\gamma_\mathrm{c})$ and $\tilde y(k)\oplus C(\mathcal{E}(P_\mathrm{c}/\gamma_\mathrm{c})\oplus\mathcal{E}(P_\mathrm{o}/\gamma_\mathrm{o}))$, respectively.
%
%
\section{Conclusions}\label{sec:conclusions}
In this work, we addressed the problem of controlling a plant described by an RNN model subject to disturbances. We proposed an LMI-based procedure for designing an output-feedback control law, consisting of a state observer and a static state-feedback controller. The design conditions are derived by introducing a novel incremental sector condition. Robustness against disturbances and estimation uncertainty is ensured by enforcing global or regional closed-loop $\delta$ISS to the RNN model. Furthermore, we showed how the design procedure can be modified to replace the static control law with a tube-based NMPC law with proven theoretical guarantees, thereby enlarging the controller’s region of attraction.\\ 
From a numerical perspective, the proposed approach exhibits a computational complexity comparable to that of established methods for linear models. The LMI-based design scales polynomially with the system dimension, making the approach applicable to a broad range of nonlinear dynamical systems. 
In practice, the dimension of a control-oriented RNN does not need to be excessively large to capture the relevant nonlinear dynamics of the system, and standard solvers perform efficiently. 
Future work will focus on extending the method to large-scale systems through distributed or decentralised implementations, leveraging decomposition-based strategies and robust controllers of the type presented in this paper to account for subsystem coupling effects. We will also explore the extension of the approach to deep RNN models.
Finally, future research will focus on defining the disturbance set for a generic disturbance model when only input–output data are available.
%
%
%
%
%

\bibliographystyle{plain}        
\bibliography{autosam}           



\appendix
\section{Appendix: Proof of the main results}~\label{appendix_A} %
In this appendix we report the proofs of the main results reported in the paper. In the following, we omit the time dependency where possible for the sake of conciseness.\smallskip\\
Before proceeding with the proofs, we need to introduce first the following lemma. 
\begin{lemma}\label{lem:constr_satisfaction}
    Consider $n_v$ vectors $v_i\in\mathcal{E}(P_i)$, where $P_i\in \mathbb{S}^{n_i}_{+}$, $i=1,\dots,n_v$. A sufficient condition for $|\sum_{i=1}^{n_v}G_iv_i|\leq b$, where $G_i$ are matrices of suitable dimensions and $b\in\R_{+}$, is 
\begin{equation}\label{eq:constr_satisfaction}
    \begin{bmatrix}
        P_1/n_v & 0 & \cdots & 0 & G_1^\top \\
        0 & P_2/n_v  & \cdots & 0 & G_2^\top \\
        \vdots & \vdots & \ddots & \vdots & \vdots \\
        0 & 0 & \cdots & P_n/n_v  & G_{n_v}^\top \\
        G_1 & G_2 & \cdots & G_{n_v} & b^2
    \end{bmatrix} \succeq 0.
\end{equation}
\hfill{}$\square$
\end{lemma}
\textbf{Proof of Lemma \ref{lem:constr_satisfaction}}.
Since  $v_i\in\mathcal{E}(P_i)$ for all $i=1,\dots,n_v$,  it follows that $\sum_{i=1}^\nu v_i^\top P_iv_i\leq n_v$, which can be rewritten in quadratic form as
\begin{equation}\label{eq:constr_satisfaction_1}
\begin{bmatrix}
    v_1^\top&\dots&v_{n_v}^\top
\end{bmatrix}
\begin{bmatrix}
        P_1/n_v & 0 & \cdots & 0\\
        0 & P_2/n_v  & \cdots & 0\\
        \vdots & \vdots & \ddots & \vdots \\
        0 & 0 & \cdots & P_n/n_v
    \end{bmatrix}
\begin{bmatrix}
    v_1\\\vdots\\v_{n_v}
\end{bmatrix}\leq 1.
\end{equation}
By applying the Schur complement, condition \eqref{eq:constr_satisfaction} is equivalent to
\[
\begin{bmatrix}
        P_1/n_v&0&\cdots &0\\
        0&P_2/n_v&\cdots&0\\
        \vdots&\vdots&\ddots&\vdots \\
        0&0 &\cdots& P_n/n_v
\end{bmatrix}\!-\!\cfrac{1}{b^2}\begin{bmatrix}
        G_1^\top\\\vdots\\G_{n_v}^\top
\end{bmatrix}
\begin{bmatrix}
        G_1&\dots&G_{n_v}
\end{bmatrix}\succeq0,
\]
which leads to
\begin{equation}\label{eq:constr_satisfaction_2}
\begin{bmatrix}
        P_1/n_v&0&\cdots &0\\
        0&P_2/n_v&\cdots&0\\
        \vdots&\vdots&\ddots&\vdots \\
        0&0 &\cdots& P_n/n_v
\end{bmatrix}\!\succeq\cfrac{1}{b^2}\begin{bmatrix}
        G_1^\top\\\vdots\\G_{n_v}^\top
\end{bmatrix}
\begin{bmatrix}
        G_1&\dots&G_{n_v}
\end{bmatrix}.
\end{equation}
From conditions \eqref{eq:constr_satisfaction_1} and \eqref{eq:constr_satisfaction_2}, we obtain
\[
\begin{bmatrix}
    v_1^\top&\dots&v_{n_v}^\top
\end{bmatrix}
\begin{bmatrix}
        G_1^\top\\\vdots\\G_{n_v}^\top
\end{bmatrix}
\begin{bmatrix}
        G_1&\dots&G_{n_v}
\end{bmatrix}
\begin{bmatrix}
    v_1\\\vdots\\v_{n_v}
\end{bmatrix}\leq b^2,
\]
which is equivalent to $|\sum_{i=1}^{n_v}G_iv_i|\leq b$, completing the proof.
\hfill{}$\square$\smallskip\\
Also, we introduce the following lemma, useful to characterize robustly positive invariant sets of $\delta$ISS systems.
\begin{lemma}\label{lem:RPI_set}
    Consider a variable $v\in\R^{n_v}$. If there exist a scalar $\gamma\in\R_+$ and matrices $P,Q_x\in\mathbb{S}_+^{n_v}$ such that $Q_x- P/\gamma\succeq0$ and, for all $v\in\mathcal{E}(P/\gamma)$, it holds
    \begin{equation}\label{eq:quadratic_dISS_function}
    \norm{v^+}^2_P-\norm{v}^2_P\leq -\norm{v}^2_{Q_x}+1,
    \end{equation}
    then $\mathcal{E}(P/\gamma)$ is forward invariant for the dynamics of $v$, i.e., $v\in\mathcal{E}(P/\gamma)$ implies $v^+\in\mathcal{E}(P/\gamma)$.\hfill{}$\square$
\end{lemma}
\textbf{Proof of Lemma \ref{lem:RPI_set}}.
Observe that $Q_x\succeq P/\gamma$
    implies $\mathcal{E}(Q_{x})\subseteq\mathcal{E}(P/\gamma)$.
    First of all, if $v\in\mathcal E(P/\gamma)\backslash\mathcal E(Q_{x})$, we have $\norm{v}_{Q_{x}}^2\geq 1$. Thus, from \eqref{eq:quadratic_dISS_function}, it follows that $\norm{v^+}^2_{P}\leq\norm{v}^2_{P}\leq\gamma$.
    On the other hand, if $v\in\mathcal E(Q_{x})$, $Q_x\succeq P/\gamma$ implies $\norm{v}^2_{Q_{x}}\geq\frac{1}{\gamma}\norm{v}^2_{P}$. Combining this with \eqref{eq:quadratic_dISS_function}, we obtain
    \begin{equation*}
        \norm{v^+}^2_{P}\leq\norm{v}^2_{P}-\frac{1}{\gamma}\norm{v}_{P}^2+1
            =\left(\cfrac{\gamma-1}{\gamma}\right)\norm{v}_{P}^2+1
            =(\gamma-1)\norm{v}_{Q_{x}}^2+1\leq\gamma, 
    \end{equation*}
    Therefore, $v^+\in\mathcal{E}(P/\gamma)$ as claimed.
\hfill{}$\square$\smallskip\\
%
%
%
%
\textbf{Proof of Lemma \ref{lem:sector_condition}.}
From the definition of function $q_i(v_i)=v_i-\sigma_i(v_i)$, we have that
\begin{equation}\label{eq:q_derivative}
\frac{\partial q_i(v_i)}{\partial v_i}=\frac{\partial}{\partial v_i}(v_i-\sigma_i(v_i))=1-\frac{\partial \sigma_i(v_i)}{\partial v_i}.
\end{equation}
In view of Assumption~\ref{ass:sigmoid_function},
\[0< \frac{\partial \sigma_i(v_i)}{\partial v_i}\leq 1,\]  
and $\partial \sigma_i(v_i)/\partial v_i=1$ only for $v_i=0$.
Consequently, from \eqref{eq:q_derivative}, the derivative of $q_i(v_i)$  with respect to $v_i$ is also bounded, giving
\begin{equation}\label{eq:q_derivative_global}
0\leq\frac{\partial q_i(v_i)}{\partial v_i}<1,
\end{equation}
and $\partial q_i(v_i)/\partial v_i=0$ only for $v_i=0$.
Condition \eqref{eq:q_derivative_global} implies that $q_i(\cdot)$ is monotonically increasing and Lipschitz continuous with unitary Lipschitz constant.

Since  $q_i(\cdot)$ is continuous and differentiable, in view of the mean value theorem, for all $(v_i,v_i+\Delta v_i)\in\R^2$, there exists a scalar $v_i^\star$ such that $v_i^\star\in[v_i,v_i+\Delta v_i]$ if $\Delta v_i\geq0$, or $v_i^\star\in[v_i+\Delta v_i,v_i]$ if $\Delta v_i<0$,
such that
\begin{equation}\label{eq:Lagrange_theorem}
\Delta q_i=q_i(v_i+\Delta v_i)-q_i(v_i)=\delta q_i(v_i^\star)\Delta v_i,
\end{equation}
where $\delta q_i(v_i^\star)=\frac{\partial{q_i(v_i)}}{\partial{v_i}}|_{v_i=v_i^\star}$.\\ 
This implies that
\begin{equation*}
    \Delta q_i(\Delta v_i-h_i\Delta q_i)=\delta q_i(v_i^\star)(1-h_i\delta q_i(v_i^\star))\Delta v_i^2.
\end{equation*}
Recall that, in view of Assumption~\ref{ass:sigmoid_function}, for $v_i^\star<0$ (resp. $v_i^\star>0$), $\frac{\partial^2 q_i(v_i)}{\partial v_i^2}<0$ (resp. $\frac{\partial^2 q_i(v_i)}{\partial v_i^2}>0$), meaning that $\delta q_i(v_i^\star)$ is strictly decreasing (resp. increasing) for $v_i^\star<0$ (resp. $v_i^\star>0$). Also, $\delta q_i(v_i^\star)\to1$ both as $v_i^\star\to+\infty$ and $v_i^\star\to-\infty$, and $\delta q_i(0)=0$.\\
Moreover, from \eqref{eq:q_derivative_global}, it holds that $0\leq\delta q_i(v_i^\star)\leq1$, $\forall v_i^\star\in\R$.
In  view of this, it follows that
\begin{itemize}
    \item for $h_i=1$, $1-h_i\delta q_i(v_i^\star)\geq0$, and $\delta q_i(v_i^\star)\geq 0$, $\forall v_i^\star\in\R$;
    \item for all $h_i>1$, $\exists\tilde v_i\in\R$ sufficiently small such that $1-h_i\delta q_i(v_i^\star)\geq 0$, i.e. $\delta q_i(v_i^\star)\leq\frac{1}{h_i}$, for any $v_i^\star\in[-\tilde v_i,\tilde v_i]$.
\end{itemize}
In particular, define
\begin{equation}\label{opt:locality_constraints}
    \begin{aligned}
    \bar v_i(h_i)=&\max_{\tilde v_i}\tilde v_i\\
    &\text{subject to}\\
    &\delta q_i(v_i^\star)\leq\frac{1}{h_i}, \forall v_i^\star\in[-\tilde v_i,\tilde v_i].
    \end{aligned}
    \end{equation} 
In view of the continuity of $\delta q_i(v_i^\star)$, $\bar v_i(h_i)$ is also continuous. Also, since $0\leq\delta q_i(v_i^\star)\leq 1$ $\forall v_i^\star$, and $\delta q_i(v_i^\star)=0$ only in $v_i^\star=0$, then $\bar v_i(h_i)\to+\infty$ as $h_i\to1^+$, and $\bar v_i(h_i)\to0^+$ as $h_i\to+\infty$.\hfill{}$\square$\smallskip\\
\textbf{Proof of Proposition \ref{prop:dISS_observer}.}
\textcolor{blue}
The proof of Proposition~\ref{prop:dISS_observer} is divided into different steps, specified here for better clarity.
\begin{itemize}
\item[1.] Prove that if \eqref{eq:dISS_condition_obs} holds and if $v(k),\hat v(k)\in\mathcal{V}(H_o)$, then
\begin{equation}\label{eq:dissipation_form_function_observer}
    \Delta V_\mathrm{o}\leq-\norm{e}^2_{Q_{\mathrm{o},x}}+\norm{ w_\mathrm{o}(k)}^2_{Q_{\mathrm{o},w_\mathrm{o}}},
    \end{equation}
    where $\Delta V_\mathrm{o}=V_\mathrm{o}(x(k+1),\hat x(k+1))-V_\mathrm{o}(x(k),\hat x(k))$, $V_\mathrm{o}(x,\hat x)=\|e\|^2_{P_\mathrm{o}}$, and $w_\mathrm{o}= [w^\top,\eta^\top]^\top$.
\item[2.] Show that, under \eqref{eq:dissipation_form_function_observer},
\eqref{eq:mRPI_set_condition_1_dISS_obs}, and \eqref{eq:mRPI_set_condition_2_dISS_obs}, if $e\in\mathcal{E}(P_\mathrm{o}/\gamma_\mathrm{o})$ and $w_\mathrm{o}\in\mathcal{E}(Q_w^0)\times\mathcal{E}(Q_\eta^0)$, then $e^+\in\mathcal{E}(P_\mathrm{o}/\gamma_\mathrm{o})$. 
\item[3.] Show that, if \eqref{eq:locality_set_dISS_obs1} and \eqref{eq:locality_set_dISS_obs2} are  also satisfied, then $v, \hat v \in \mathcal V(H_\mathrm{o})$ for all $v, \hat v \in \mathbb R^\nu$ in the case $\mathcal I(H_\mathrm{o}) = \emptyset$, and for all $(\hat x, u)$ satisfying \eqref{eq:locality_condition_obs} in the case $\mathcal I(H_\mathrm{o}) \neq \emptyset$.
\item[4.] Conclude that~\eqref{eq:observer_dynamics} has ISS observer error dynamics.  
\end{itemize}
In the following, the proofs of steps 1-4 are provided.\smallskip\\
1. In view of \eqref{eq:sector_inequality} and Lemma \ref{lem:sector_condition}, for any $H_\mathrm{o}\in\mathbb{D^\nu_+}$ with $H_\mathrm{o}\succeq I_\nu$, if $v,\hat{v}\in \mathcal{V}(H_\mathrm{o})$ then, for all $S_\mathrm{o}\in\mathbb{D^\nu_+}$
\begin{equation}\label{eq:sector_inequality_obs1}
    \Delta q(v,\hat{v})^\top S_\mathrm{o}(\Delta v-H_\mathrm{o}\Delta q(v,\hat{v}))\geq 0
\end{equation}
where $\Delta v=v-\hat{v}=(\tilde{A}-\tilde{L}C)e+\tilde{D}w-\tilde{L}\eta$. For compactness we denote $\Delta q_o=\Delta q(v,\hat{v})$ and $\phi_o=[e^\top,\Delta q_o^\top,w_\mathrm{o}^\top]^\top$ and we rewrite \eqref{eq:sector_inequality_obs1} as
\begin{equation}\label{eq:sector_inequality_obs2}
    \Delta q_o^\top S_\mathrm{o}((\tilde{A}-\tilde{L}C)e+\begin{bmatrix}
      \tilde{D}&  -\tilde{L}
    \end{bmatrix}w_\mathrm{o}-H_\mathrm{o}\Delta q_o)=
    \Delta q_o^\top S_\mathrm{o}(\tilde{A}_{\mathrm{L}}e+\tilde{D}_{\mathrm{o}}w_\mathrm{o}-H_\mathrm{o}\Delta q_o)\geq 0
\end{equation}
where $\tilde{A}_{\mathrm{L}}=\tilde{A}-\tilde{L}C$ and $\tilde{D}_{\mathrm{o}}=\begin{bmatrix}
      \tilde{D}&  -\tilde{L}
    \end{bmatrix}$. The latter can be rewritten as
\begin{equation}\label{eq:sector_condition_quadratic_obs}
    \phi_o^\top\begin{bmatrix}
    0& \tilde{A}_{\mathrm{L}}^\top S_\mathrm{o} &0\\
    S_\mathrm{o}\tilde{A}_{\mathrm{L}}&-2S_\mathrm{o}H_\mathrm{o}& S_\mathrm{o}\tilde D_{\mathrm{o}}\\
    0&\tilde D_{\mathrm{o}}^\top S_\mathrm{o}&0
\end{bmatrix}\phi_o\geq0.
\end{equation}
 Defining $A_{\mathrm{L}}=A-LC$, the dynamics of $e$ is  \begin{equation}\label{eq:error_dynamics_observer}
    e^+=x^+-\hat{x}^+={A}_{\mathrm{L}}e+D w-L \eta+B_q \Delta q_o
    =\begin{bmatrix}
        {A}_{\mathrm{L}}&B_q&D_{\mathrm{o}}
    \end{bmatrix}\phi_o,
    \end{equation}
    where $D_{\mathrm{o}}=\begin{bmatrix}D&L\end{bmatrix}$. We can write
\begin{equation}\label{eq:Lyapunov_function_increment_observer}
\begin{aligned}
\Delta V_\mathrm{o}
&=V_\mathrm{o}(x(k+1),\hat x(k+1))-V_\mathrm{o}(x(k),\hat x(k))\\
&=\phi_o^\top\left(\begin{bmatrix}
        {A}_{\mathrm{L}}^\top\\B_q^\top\\D_{\mathrm{o}}^\top
    \end{bmatrix}P_\mathrm{o}\begin{bmatrix}
        {A}_{\mathrm{L}}&B_q&D_{\mathrm{o}}
    \end{bmatrix}-\begin{bmatrix}
        P_\mathrm{o}&0&0\\0&0&0
        \\0&0&0
    \end{bmatrix}\right)\phi_o.
\end{aligned}
\end{equation}
From \eqref{eq:sector_condition_quadratic_obs} and \eqref{eq:error_dynamics_observer}, if $v,\hat{v}\in \mathcal{V}(H_\mathrm{o})$ it is possible to guarantee \eqref{eq:dissipation_form_function_observer} 
and, in view of Theorem \ref{th:local_dISS}, that the observer error dynamics are $\delta$ISS, by imposing
\begin{equation}\label{eq:dissipation_form_function_observer_2}
    \Delta V_\mathrm{o}+\phi_o^\top\begin{bmatrix}
    0&\tilde{A}_{\mathrm{L}}^\top S_\mathrm{o} &0\\
    S_\mathrm{o}\tilde{A}_{\mathrm{L}}&-2S_\mathrm{o}H_\mathrm{o}& S_\mathrm{o}\tilde D_{\mathrm{o}}
    0&\tilde D_{\mathrm{o}}^\top S_\mathrm{o}&0
\end{bmatrix}\phi_o\leq-\norm{e}^2_{Q_{\mathrm{o},x}}+\norm{w_\mathrm{o}}^2_{Q_{o,w_\mathrm{o}}}\,.
\end{equation}
Using \eqref{eq:Lyapunov_function_increment_observer}, condition \eqref{eq:dissipation_form_function_observer_2} is satisfied by imposing
\begin{equation}\label{eq:dISS_condition_2_obs_zero}
\begin{bmatrix}
        {A}_{\mathrm{L}}^\top\\B_q^\top\\D_{\mathrm{o}}^\top
    \end{bmatrix}P_\mathrm{o}\begin{bmatrix}
        {A}_{\mathrm{L}}&B_q&D_{\mathrm{o}}
    \end{bmatrix}-
\begin{bmatrix}
    P_\mathrm{o}-Q_{\mathrm{o},x}&-\tilde A_{\mathrm{L}}^\top S_\mathrm{o} &0\\
    -S_\mathrm{o}\tilde A_{\mathrm{L}}&2S_\mathrm{o}H_\mathrm{o}& -S_\mathrm{o}\tilde D_{\mathrm{o}}\\
    0&-\tilde D_{\mathrm{o}}^\top S_\mathrm{o}&Q_{o,w_\mathrm{o}}
    \end{bmatrix}
    \preceq 0
\end{equation}
which, in view of the Schur complement, is equivalent to
\begin{equation}\label{eq:dISS_condition_2_obs}
\begin{bmatrix}
    P_\mathrm{o}-Q_{\mathrm{o},x}&-\tilde A_{\mathrm{L}}^\top S_\mathrm{o} &0&A_{\mathrm{L}}^\top P_\mathrm{o}\\
    -S_\mathrm{o}\tilde A_{\mathrm{L}}&2S_\mathrm{o}H_\mathrm{o}& -S_\mathrm{o}\tilde D_{\mathrm{o}}&B_q^\top P_\mathrm{o}\\
    0&-\tilde D_{\mathrm{o}}^\top S_\mathrm{o}&Q_{o,w_\mathrm{o}}&D_{\mathrm{o}}^\top P_\mathrm{o}\\
    P_\mathrm{o}A_{\mathrm{L}}&P_\mathrm{o}B_q&P_\mathrm{o}D_{\mathrm{o}}&P_\mathrm{o}
    \end{bmatrix}
    \succeq 0.
\end{equation}
Substituting $N=P_\mathrm{o}L$ and $\tilde N=S_\mathrm{o}\tilde L$ into \eqref{eq:dISS_condition_2_obs}, yields condition \eqref{eq:dISS_condition_obs}.\smallskip\\
2. In view of Assumptions~\ref{ass:disturbance} and \ref{ass:output_noise}, condition \eqref{eq:mRPI_set_condition_1_dISS_obs} implies \[\norm{w_\mathrm{o}}^2_{Q_{o,w_\mathrm{o}}}=\norm{\begin{bmatrix}
    w\\\eta
\end{bmatrix}}^2_{Q_{o,w_\mathrm{o}}}\leq \frac{1}{2}\left(\norm{w}^2_{Q_w^0}+\norm{\eta}^2_{Q_\eta^0}\right)\leq 1.\] 
From \eqref{eq:dissipation_form_function_observer}, this implies that 
\begin{equation}  
\|e^+\|^2_{P_\mathrm{o}}-\|e\|^2_{P_\mathrm{o}}\leq-\norm{e}^2_{Q_{\mathrm{o},x}}+1.\label{eq:dissipation_form_function_4_obs}\end{equation}
By Lemma~\ref{lem:RPI_set}, since 
$e$ satisfies \eqref{eq:dissipation_form_function_4_obs} and \eqref{eq:mRPI_set_condition_2_dISS_obs} holds,  $e\in\mathcal E(P_\mathrm{o}/\gamma_\mathrm{o})$ implies $e^+\in\mathcal E(P_\mathrm{o}/\gamma_\mathrm{o})$ as claimed.\smallskip\\
%
3. First note that, in the trivial case where $\mathcal{I}(H_\mathrm{o})=\emptyset$, the diagonal elements of $H_\mathrm{o}$ are $h_{\mathrm{o},i}=1$ for all $i=1,\dots,\nu$ by definition and, since $\bar{v}_i(h_{\mathrm{o},i})\rightarrow +\infty$ as $h_{\mathrm{o},i}\rightarrow 1^+$ for all $i=1,\dots,\nu$, then $\mathcal{V}(H_\mathrm{o})$ is unbounded and $v,\hat{v}\in \mathcal{V}(H_\mathrm{o})$ for all $v,\hat{v}\in\mathbb{R}^{\nu}$.\\
Now, let us consider the case $\mathcal{I}(H_\mathrm{o})\neq\emptyset$. Note that $v=\tilde A_{\mathrm{L}}x+\tilde Bu+\tilde LCx+\tilde Dw=\tilde A\hat x+\tilde B u+\tilde Dw+\tilde Ae$ and $\hat v=\tilde A_{\mathrm{L}}\hat x+\tilde Bu+\tilde LCx+\tilde L\eta=\tilde A\hat x+\tilde B u+\tilde LCe+\tilde L\eta$.
If $w$ satisfies Assumption~\ref{ass:disturbance} and $e\in\mathcal{E}(P_\mathrm{o}/\gamma_\mathrm{o})$, a sufficient condition for $v\in\mathcal V(H_\mathrm{o})$ is that $\tilde A\hat x+\tilde B u\in\mathcal V(H_\mathrm{o})\ominus\tilde D\mathcal{E}(Q_w^0)\ominus\tilde A\mathcal{E}(P_\mathrm{o}/\gamma_\mathrm{o})$. The latter set exists only if $\tilde D\mathcal E(Q_w^0)\oplus\tilde A\mathcal E(P_\mathrm{o}/\gamma_\mathrm{o})\subseteq\mathcal V(H_\mathrm{o})$, i.e., only if
for all $i\in\mathcal I(H_\mathrm{o})$ and for all $e\in\mathcal{E}(P_\mathrm{o}/\gamma_\mathrm{o})$, $w\in \mathcal{E}(Q_w^0)$ it holds that
$|\tilde A_{i} e+ \tilde D_{i} w|\leq \bar v_i(h_{\mathrm{o},i})$. By Lemma~\ref{lem:constr_satisfaction}, this condition is guaranteed by \eqref{eq:locality_set_dISS_obs1}.

Similarly, under Assumption~\ref{ass:output_noise}, $\hat{v}\in\mathcal V(H_\mathrm{o})$ if $A\hat x+\tilde B u\in\mathcal V(H_\mathrm{o})\ominus\tilde LC\mathcal{E}(P_\mathrm{o}/\gamma_\mathrm{o})\ominus\tilde L\mathcal{E}(Q_\eta^0)$. The latter set exists only if $(e,\eta)\in\mathcal E(P_\mathrm{o}/\gamma_\mathrm{o})\times\mathcal E(Q_\eta^0)$ implies $\tilde LCe+\tilde L\eta\in\mathcal V(H_\mathrm{o})$, i.e.,
$|\tilde L_{i}C e+ \tilde L_{i} \eta|\leq \bar v_i(h_{\mathrm{o},i})$, for all $i\in\mathcal I(H_\mathrm{o})$,
which, according to Lemma~\ref{lem:constr_satisfaction}, holds if
\begin{equation}\label{eq:locality_set_dISS_obs2_2}
    \begin{bmatrix}
        P_\mathrm{o}/(2\gamma_\mathrm{o}) & 0 & C^\top\tilde L_{i}^\top\\ 
        0 & Q_\eta^0/2 & \tilde L_{i}^\top\\
        \tilde L_{i}C&\tilde L_{i} & v_i(h_{\mathrm{o},i})^2 
    \end{bmatrix}\succeq 0, \forall i\in\mathcal{I}(H_\mathrm{o}).
\end{equation}
By congruence transformation, condition \eqref{eq:locality_set_dISS_obs2_2} is equivalent to
\begin{equation*}
    \begin{bmatrix}
        P_\mathrm{o}/(2\gamma_\mathrm{o}) & 0 & C^\top\tilde L_{i}^\top s_{\mathrm{o},i}\\ 
        0 & Q_\eta^0/2 & \tilde L_{i}^\top s_{\mathrm{o},i}\\
        s_{\mathrm{o},i}\tilde L_{i}C&s_{\mathrm{o},i}\tilde L_{i} & v_i(h_{\mathrm{o},i})^2s_{\mathrm{o},i}^2 
    \end{bmatrix}\succeq 0, \forall i\in\mathcal{I}(H_\mathrm{o}),
\end{equation*}
that, recalling that $\tilde N=\tilde LS_{\mathrm{o}}$, is equivalent to \eqref{eq:locality_set_dISS_obs2}.
\smallskip\\
4. In case $\mathcal I(H_\mathrm{o}) =\emptyset$, the Lyapunov function $V_\mathrm{o}(x,\hat x)$ satisfies~\eqref{eq:dissipation_form_function_observer} for all $x, \hat x\in\R^n$, $w\in\mathcal{E}(Q_w^0)$, and $\eta\in\mathcal{E}(Q_\eta^0)$. Following the same arguments as in the proof of Theorem~2 in~\cite{magni2006regional}, \eqref{eq:dissipation_form_function_observer} implies the existence of functions $\beta_{\mathrm{o}}\in\mathcal{KL}$, $\gamma^w_\mathrm{o}\in\mathcal{K}$, and $\gamma^\eta_o\in\mathcal{K}$ such that for any $k\in\mathbb{Z}_{\geq0}$, any initial estimation error $e(0)\in\R^n$ and any pair of disturbance and noise $w\in\mathcal{E}(Q^0_w)$ and $\eta\in\mathcal{E}(Q^0_\eta)$, it holds that
\begin{equation*}
    \norm{e(k)}\leq\beta_{\mathrm{o}}(\norm{e(0)},k)       +\gamma_\mathrm{o}^w(\max_{h\geq 0}\norm{w(h)})+\gamma_\mathrm{o}^\eta(\max_{h\geq 0}\norm{\eta(h)}).
\end{equation*}
Therefore, \eqref{eq:observer_dynamics} has ISS observer error dynamics over the sets $\R^n$, $\mathcal{E}(Q_w^0)$, and $\mathcal{E}(Q_\eta^0)$ according to Definition~\ref{def:delta_ISS_observer}.\\
Conversely, if $\mathcal{I}(H_\mathrm{o}) \neq \emptyset$, then, by invariance of the set $\mathcal{E}(P_\mathrm{o}/\gamma_\mathrm{o})$, any initial condition $e(0)\in\mathcal{E}(P_\mathrm{o}/\gamma_\mathrm{o})$ implies that $e(k)\in\mathcal{E}(P_\mathrm{o}/\gamma_\mathrm{o})$ for all $k\in\mathbb{Z}_{\geq 0}$. Moreover, along the resulting trajectories, the Lyapunov function satisfies~\eqref{eq:dissipation_form_function_observer} for all $k\in\mathbb{Z}_{\geq 0}$.
Using analogous arguments as above, this implies that system~\eqref{eq:observer_dynamics} has ISS observer error dynamics with respect to the sets $\mathcal{E}(P_\mathrm{o}/\gamma_\mathrm{o})$, $\mathcal{E}(Q_{w}^0)$, and $\mathcal{E}(Q_\eta^0)$, concluding the proof.
\hfill{}$\square$\smallskip\\
\textbf{Proof of Lemma 
\ref{lem:NScond_obs}.}
To prove that the feasibility of~\eqref{eq:dISS_condition_obs} implies the detectability of $(A,C)$, observe that, since $N=P_\mathrm{o}L$ and $\tilde N=S_\mathrm{o}\tilde L$, inequality \eqref{eq:dISS_condition_obs} is equivalent to \eqref{eq:dISS_condition_2_obs}. Defining
\begin{equation}\label{eq:congr_transf}
T=\begin{bmatrix}
        I&0&0&0\\
        0&0&I&0\\
        0&0&0&I\\
        0&I&0&0
    \end{bmatrix},
\end{equation}
it follows from \eqref{eq:dISS_condition_2_obs} that
\begin{equation}\label{eq:dISS_condition_3_obs}
T^\top\begin{bmatrix}
    P_\mathrm{o}-Q_{\mathrm{o},x}&-\tilde A_{\mathrm{L}}^\top S_\mathrm{o} &0&A_{\mathrm{L}}^\top P_\mathrm{o}\\
    -S_\mathrm{o}\tilde A_{\mathrm{L}}&2S_\mathrm{o}H_\mathrm{o}& -S_\mathrm{o}\tilde D_{\mathrm{o}}&B_q^\top P_\mathrm{o}\\
    0&-\tilde D_{\mathrm{o}}^\top S_\mathrm{o}&Q_{o,w_\mathrm{o}}&D_{\mathrm{o}}^\top P_\mathrm{o}\\
    P_\mathrm{o}A_{\mathrm{L}}&P_\mathrm{o}B_q&P_\mathrm{o}D_{\mathrm{o}}&P_\mathrm{o}
    \end{bmatrix}T=\begin{bmatrix}
    P_\mathrm{o}-Q_{\mathrm{o},x}&A_{\mathrm{L}}^\top P_\mathrm{o}&-\tilde A_{\mathrm{L}}^\top S_\mathrm{o} &0\\
    P_\mathrm{o}A_{\mathrm{L}}&P_\mathrm{o}&P_\mathrm{o}B_q&P_\mathrm{o}D_{\mathrm{o}}\\
    -S_\mathrm{o}\tilde A_{\mathrm{L}}&B_q^\top P_\mathrm{o}&2S_\mathrm{o}H_\mathrm{o}& -S_\mathrm{o}\tilde D_{\mathrm{o}}\\
    0&D_{\mathrm{o}}^\top P_\mathrm{o}&-\tilde D_{\mathrm{o}}^\top S_\mathrm{o}&Q_{o,w_\mathrm{o}}
    \end{bmatrix}
    \succeq 0.
\end{equation}
Note that a necessary condition for \eqref{eq:dISS_condition_3_obs} is 
\begin{equation*}
    \begin{bmatrix}
    P_\mathrm{o}-Q_{\mathrm{o},x}&A_{\mathrm{L}}^\top P_\mathrm{o}\\
    P_\mathrm{o}A_{\mathrm{L}}&P_\mathrm{o}
    \end{bmatrix}\succeq0
\end{equation*}
which, given that $P_\mathrm{o}$ is invertible by design, can be rewritten using the Schur complement as
$P_\mathrm{o}-Q_{\mathrm{o},x}-A_{\mathrm{L}}^\top P_\mathrm{o}A_{\mathrm{L}}\succeq 0$. This inequality holds if $A_{\mathrm{L}}$ is a Schur matrix having quadratic Lyapunov function $V_\mathrm{o}(x)=x^\top P_\mathrm{o}x$. This is possible only if $(A,C)$ is detectable.\smallskip\\
We now prove that the detectability of $(A,C)$ implies the feasibility of~\eqref{eq:dISS_condition_obs}. Leveraging the Schur complement, note that condition \eqref{eq:dISS_condition_3_obs} can be rewritten as
\begin{equation}\label{eq:dISS_condition_4_obs} 
    \begin{bmatrix}
    P_\mathrm{o}-Q_{\mathrm{o},x}&A_{\mathrm{L}}^\top P_\mathrm{o}\\
    P_\mathrm{o}A_{\mathrm{L}}&P_\mathrm{o}
    \end{bmatrix}-M_\mathrm{o}Q_{H_\mathrm{o}w_\mathrm{o}}^{-1}M_\mathrm{o}^\top\succeq0
\end{equation}
where
\begin{equation*}
M_\mathrm{o}=\begin{bmatrix}
    -\tilde A_{\mathrm{L}}^\top S_\mathrm{o}&0\\
    P_\mathrm{o}B_q&P_\mathrm{o}D_{\mathrm{o}}
\end{bmatrix}, \ \ Q_{H_\mathrm{o}w_\mathrm{o}}=\begin{bmatrix}
    2S_\mathrm{o}H_\mathrm{o}&-S_\mathrm{o}\tilde D\\
    -\tilde D^\top S_\mathrm{o}&Q_{o,w_\mathrm{o}}
\end{bmatrix}. 
\end{equation*}
Since $(A,C)$ is detectable, there exists $L$ such that $A_{\mathrm{L}}=A-LC$ is Schur stable. In view of the Schureness of $A_{\mathrm{L}}$, we can select $P_\mathrm{o}$ and $Q_{\mathrm{o},x}$ such that 
\begin{equation*}
    \begin{bmatrix}
    P_\mathrm{o}-Q_{\mathrm{o},x}&A_{\mathrm{L}}^\top P_\mathrm{o}\\
    P_\mathrm{o}A_{\mathrm{L}}&P_\mathrm{o}
    \end{bmatrix}\succeq \bar lI,
\end{equation*}
where $\bar l\in\R_+$. This allows us to fulfill \eqref{eq:dISS_condition_4_obs} by setting
\begin{equation}\label{eq:dISS_condition_5_obs}
    \bar lI-M_\mathrm{o} Q_{H_\mathrm{o}w_\mathrm{o}}^{-1}M_\mathrm{o}^\top\succeq0.
\end{equation}
Condition \eqref{eq:dISS_condition_5_obs} is satisfied if, and only if, there exist matrices $S_\mathrm{o}$, $H_\mathrm{o}$ and $Q_{o,{w_\mathrm{o}}}$ such that $\norm{M_\mathrm{o} Q_{H_\mathrm{o}w_\mathrm{o}}^{-1} M_\mathrm{o}^\top} \leq \bar l$. Furthermore, being $Q_{o,{w_\mathrm{o}}}$ symmetric, then $Q_{H_\mathrm{o}w_\mathrm{o}}$ is also symmetric and has real eigenvalues. It follows that
\[\norm{M_\mathrm{o}Q_{H_\mathrm{o}w_\mathrm{o}}^{-1}M_\mathrm{o}^\top}\leq \norm{M_\mathrm{o}}^2\norm{Q_{H_\mathrm{o}w_\mathrm{o}}^{-1}}\leq\cfrac{\norm{M_\mathrm{o}}^2}{|\lambda_{\text{min}}(Q_{H_\mathrm{o}w_\mathrm{o}})|}.\]
Therefore, condition \eqref{eq:dISS_condition_5_obs} can be satisfied by imposing
\begin{equation}\label{eq:sufficient_condition_obs}
    |\lambda_{\text{min}}(Q_{H_\mathrm{o}w_\mathrm{o}})|\geq\cfrac{\norm{M_\mathrm{o}}^2}{\bar l}.
\end{equation}
According to the Geršgorin theorem~\cite{Varga2004Gersgorin}, all eigenvalues of the matrix $Q_{H_\mathrm{o}w_\mathrm{o}} = [q_{ij}] \in \mathbb{R}^{\nu+d+p\times\nu+d+p}$ lie within the union of Geršgorin discs, which are centred at $q_{ii}$ with radii $r_i = \sum_{j \neq i} |q_{ij}|$, for all $i = 1, \dots, \nu+d+p$. By choosing, for instance, $S_\mathrm{o} = \frac{1}{2} I_\nu$, $Q_{o,{w_\mathrm{o}}} = \alpha I_{d+p}$ and $H_\mathrm{o} = \alpha I_\nu$, where $\alpha\in\R_{\geq 0}$, we have $q_{ii} = \alpha$ for all $i=1,\dots,\nu+d$. Moreover, note that $\bar l$, $M_\mathrm{o}$, and $r_i$ for $i = 1,\dots,\nu+d$ do not depend on $\alpha$, and
\[\lambda_{\text{min}}(Q_{H_\mathrm{o}w_\mathrm{o}})\geq\alpha-\max_{i=1,\dots,\nu+d}r_i.\] 
Therefore, increasing $\alpha$ effectively shifts the centres of the Geršgorin discs toward larger values along the real axis in the complex plane, increasing the eigenvalues of $Q_{H_\mathrm{o}w_\mathrm{o}}$. Thus, condition \eqref{eq:sufficient_condition_obs}
can be satisfied by selecting $\alpha$ such that \[\alpha\geq\cfrac{\norm{M_\mathrm{o}}^2}{\bar l}+\max_{i=1,\dots,\nu+d}r_i,\] 
thus completing the proof.
\hfill{}$\square$\smallskip\\
\textbf{Proof of Proposition \ref{prop:dISS}.}
The proof of Proposition~\ref{prop:dISS} is divided in different steps, specified here for better clarity.
\begin{itemize}
\item[1.] Prove that if \eqref{eq:dISS_condition} holds and if, for each $j=1,2$, $\hat v_j(k)=\tilde{A}_K\hat x_j(k)+\tilde{B}(\bar{u}-K\bar{x})+\tilde{D}_\mathrm{c}w_{\mathrm{c},j}(k)\in\mathcal{V}(H_\mathrm{c})$, then
\begin{equation}\label{eq:dissipation_form_function}
    V(\hat x_1(k+1),\hat x_2(k+1))-V(\hat x_1(k),\hat x_2(k))\leq-\norm{\Delta \hat x(k)}^2_{Q_{\mathrm{c},x}}+\norm{\Delta w_\mathrm{c}(k)}^2_{Q_{c,w_\mathrm{c}}},
    \end{equation}
    where $V(\hat x_1,\hat x_2)=\|\Delta \hat x\|^2_{P_\mathrm{c}}$, $\Delta \hat x=\hat x_1-\hat x_2$, $Q_{\mathrm{c},x}\in\mathbb{S}_+^n$, and $\Delta w_\mathrm{c}=w_{\mathrm{c},1}-w_{\mathrm{c},2}$.
\item[2.] Prove that, if also \eqref{eq:mRPI_set_condition_1_dISS}-\eqref{eq:locality_eq_dISS} hold, then for each $j=1,2$, if $\hat x_j(k)\in\mathcal{E}(P_\mathrm{c}/\gamma_\mathrm{c})\oplus\bar{x}$ then $\hat x_j(k+1)\in\mathcal{E}(P_\mathrm{c}/\gamma_\mathrm{c})\oplus\bar{x}$ for all $w_{\mathrm{c},j}(k)\in\mathcal{E}(Q^0_\eta)\times\mathcal{E}(P_\mathrm{o}/\gamma_\mathrm{o})$ and $\hat v_j(k)\in \mathcal{V}(H_\mathrm{c})$.
\item[3.] Prove that if \eqref{eq:locality_set_dISS_ctrobs1}-\eqref{eq:locality_eq_dISS_ctrobs} are satisfied, then \eqref{eq:locality_condition_obs} holds for all $\hat x\in\mathcal{E}(P_\mathrm{c}/\gamma_\mathrm{c})\oplus\bar x$.
\item[4.] Conclude that $V(\hat x_1,\hat x_2)$ serves as a dissipation-form Lyapunov function. 
\end{itemize}
In the following, the proofs of steps 1-4 are provided.\smallskip\\
1. For each $j=1,2$, consider $\hat x_j\in\R^n$ and $w_{\mathrm{c},j}\in\mathcal E(Q_{w_\mathrm{c}}^0)$, and denote $\hat x_j^+=A_\mathrm{K}\hat x_j+B(\bar u -K\bar x)+D_{\mathrm{c}}w_{\mathrm{c},j}+B_qq(\hat v_j)$. 
The dynamics of $\Delta \hat x$ is  \begin{equation}\label{eq:error_dynamics}
    \Delta \hat x^+=\hat x_1^+-\hat x_2^+=A_\mathrm{K}\Delta \hat x+D_{\mathrm{c}}\Delta w_\mathrm{c}+B_q (q(\hat v_1)-q(\hat v_2)).
\end{equation}
    Noting that $\Delta \hat v=\hat v_1-\hat v_2=\tilde A_\mathrm{K}\Delta \hat x+\tilde D\Delta w_\mathrm{c}$, in view of \eqref{eq:sector_inequality}, for all $S_\mathrm{c}\in\mathbb D_+^\nu$, the nonlinear function $\Delta q(\hat v_1,\hat v_2)$ satisfies, for all $\hat v_1,\hat v_2\in\mathcal{V}(H_\mathrm{c})$,
\begin{equation}\label{eq:sector_condition}
\Delta q(\hat v_1,\hat v_2)^\top S_\mathrm{c}(\tilde A_\mathrm{K}\Delta \hat x+\tilde D_{\mathrm{c}}\Delta w_\mathrm{c}-H_\mathrm{c}\Delta q(\hat v_1,\hat v_2))\geq 0
\end{equation}
which, letting $\phi=\begin{bmatrix}\Delta \hat x^\top,&\Delta q(\hat v_1,\hat v_2)^\top, &\Delta w_\mathrm{c}^\top\end{bmatrix}^\top$
is equivalent to 
\begin{equation}\label{eq:sector_condition_quadratic}
    \phi^\top\begin{bmatrix}
    0&\tilde A_\mathrm{K}^\top S_\mathrm{c} &0\\
    S_\mathrm{c}\tilde A_\mathrm{K}&-2S_{\mathrm{c}}H_\mathrm{c}& S_\mathrm{c}\tilde D_{\mathrm{c}}\\
    0&\tilde D_{\mathrm{c}}^\top S_\mathrm{c}&0
\end{bmatrix}\phi\geq0.
\end{equation}
Given that $V(\hat x_1,\hat x_2)=\norm{\Delta \hat x}_{P_\mathrm{c}}^2$, where $P_\mathrm{c}\in\mathbb{S}^n_+$, we can write
\begin{equation}\label{eq:Lyapunov_function_increment}
\begin{aligned}
\Delta V&=(A_\mathrm{K}\Delta \hat x+D_{\mathrm{c}}\Delta w_\mathrm{c}+B_q\Delta q(\hat v_1,\hat v_2))^\top P_\mathrm{c}(A_\mathrm{K}\Delta \hat x+D_{\mathrm{c}}\Delta w_\mathrm{c} +B_q\Delta q(\hat v_1,\hat v_2))-\Delta \hat x^\top P_\mathrm{c}\Delta \hat x\\
&=\phi^\top\left(\begin{bmatrix}
        A_\mathrm{K}^\top\\B_q^\top\\D_{\mathrm{c}}^\top
    \end{bmatrix}P_\mathrm{c}\begin{bmatrix}
        A_\mathrm{K}&B_q&D_{\mathrm{c}}
    \end{bmatrix}-\begin{bmatrix}
        P_\mathrm{c}&0&0\\0&0&0
        \\0&0&0
    \end{bmatrix}\right)\phi.
\end{aligned}
\end{equation}
Therefore, it is possible to guarantee \eqref{eq:dissipation_form_function}, for all $\hat v_1,\hat v_2\in\mathcal{V}(H_\mathrm{c})$, under \eqref{eq:sector_condition_quadratic} by imposing
\begin{equation}\label{eq:dissipation_form_function_2}
    \Delta V+\phi^\top\begin{bmatrix}
    0&\tilde A_\mathrm{K}^\top S_\mathrm{c} &0\\
    S_\mathrm{c}\tilde A_\mathrm{K}&-2S_\mathrm{c}H_\mathrm{c}& S_\mathrm{c}\tilde D_{\mathrm{c}}\\
    0&\tilde D_{\mathrm{c}}^\top S_\mathrm{c}&0
\end{bmatrix}\phi\leq-\norm{\Delta \hat x}^2_{Q_{\mathrm{c},x}}+\norm{\Delta \hat w}^2_{Q_{c,w_\mathrm{c}}}\,.
\end{equation}
Following similar steps as in \eqref{eq:dissipation_form_function_observer_2}-\eqref{eq:dISS_condition_2_obs}, we derive that a sufficient condition for \eqref{eq:dissipation_form_function_2} is
\begin{equation}\label{eq:dISS_condition_P}
\begin{bmatrix}
    P_\mathrm{c}-Q_{\mathrm{c},x}&-\tilde A_\mathrm{K}^\top S_\mathrm{c} &0&A_\mathrm{K}^\top P_\mathrm{c}\\
    -S_\mathrm{c}\tilde A_\mathrm{K}&2S_\mathrm{c}H_\mathrm{c}& -S_\mathrm{c}\tilde D_{\mathrm{c}}&B_q^\top P_\mathrm{c}\\
    0&-\tilde D_{\mathrm{c}}^\top S_\mathrm{c}&Q_{c,w_\mathrm{c}}&D_{\mathrm{c}}^\top P_\mathrm{c}\\
    P_\mathrm{c}A_\mathrm{K}&P_\mathrm{c}B_q&P_\mathrm{c}D_{\mathrm{c}}&P_\mathrm{c}
    \end{bmatrix}
    \succeq 0.
\end{equation}
Define matrices $Q_{\mathrm{c}}=P_\mathrm{c}^{-1}$, $U_\mathrm{c}=S_\mathrm{c}^{-1}$, $\tilde Q_{\mathrm{c},x}=Q_\mathrm{c}Q_{\mathrm{c},x}Q_{\mathrm{c}}$, and $M=\diag(Q_{\mathrm{c}},U_\mathrm{c},I_d,Q_{\mathrm{c}})$. By applying a congruence transformation~\cite{boyd1994linear}, condition \eqref{eq:dISS_condition_P} is equivalent to
\begin{equation}\label{eq:dISS_condition_congr}
M\begin{bmatrix}
    P_\mathrm{c}-Q_{\mathrm{c},x}&-\tilde A_\mathrm{K}^\top S_\mathrm{c} &0&A_\mathrm{K}^\top P_\mathrm{c}\\
    -S_\mathrm{c}\tilde A_\mathrm{K}&2S_\mathrm{c}H_\mathrm{c}& -S_\mathrm{c}\tilde D_{\mathrm{c}}&B_q^\top P_\mathrm{c}\\
    0&-\tilde D_{\mathrm{c}}^\top S_\mathrm{c}&Q_{c,w_\mathrm{c}}&D_{\mathrm{c}}^\top P_\mathrm{c}\\
    P_\mathrm{c}A_\mathrm{K}&P_\mathrm{c}B_q&P_\mathrm{c}D_{\mathrm{c}}&P_\mathrm{c}
\end{bmatrix}M^\top\\
=\begin{bmatrix}
    Q_{\mathrm{c}}-\tilde Q_{\mathrm{c},x}&-Q_{\mathrm{c}}\tilde A_\mathrm{K}^\top &0&Q_\mathrm{c}A_\mathrm{K}^\top\\
    -\tilde A_\mathrm{K}Q_\mathrm{c}&2H_\mathrm{c}U_\mathrm{c}& -\tilde D_{\mathrm{c}}&U_\mathrm{c}B_q^\top\\
    0&-\tilde D_{\mathrm{c}}^\top &Q_{c,w_\mathrm{c}}&D_{\mathrm{c}}^\top\\
    A_\mathrm{K}Q_\mathrm{c}&B_qU_\mathrm{c}&D_{\mathrm{c}}&Q_{\mathrm{c}}
\end{bmatrix}\succeq 0,
\end{equation}
which is equivalent to \eqref{eq:dISS_condition} by setting $Z=KQ_\mathrm{c}$.
\smallskip\\
2. First note that \eqref{eq:locality_eq_dISS} implies that the state and input equilibrium pair $(\bar x,\bar u)$ satisfies $\bar{v}=\tilde A\bar x+\tilde B\bar u\in\mathcal{V}(H_\mathrm{c})$. Now assume that, for each $j=1,2$, $\hat x_j\in\mathcal{E}(P_\mathrm{c}/\gamma_\mathrm{c})\oplus \bar{x}$ and $w_{\mathrm{c},j}\in \mathcal{E}(Q_{w_\mathrm{c}}^0)$. \\
First, we show that this implies that $\hat v_j\in\mathcal V(H_\mathrm{c})$. 
    %
    In fact, left- and right-multiplying \eqref{eq:locality_set_dISS} by  $\diag{(P_\mathrm{c},I_{p+n+1})}$ and setting $K=ZQ_\mathrm{c}^{-1}$, yields 
    \begin{equation*}
\begin{bmatrix}\
        P_\mathrm{c}/(2\gamma_\mathrm{c})\!\!&\!\!0\!\!&\!\!\tilde A_{\mathrm{K},i}^\top\\
        0\!\!&\!\!Q_{w_\mathrm{c}}^0/2\!\!&\!\!\tilde D_{\mathrm{c},i}\\
        \tilde A_{\mathrm{K},i}\!\!&\!\!\tilde D_{\mathrm{c},i}\!\!&\!\!(\bar v_i(h_{\mathrm{c},i})\!-\!|\tilde A_i\bar x\!+\!\tilde B_i\bar u|)^2
        \end{bmatrix}\!\!\succeq 0,
\end{equation*}
for all $i\in\mathcal{I}(H_\mathrm{c})$, which, by Lemma~\ref{lem:constr_satisfaction}, provides a sufficient condition such that 
    \[|\tilde A_{\mathrm{K},i} (\hat x_j-\bar x)+ \tilde D_{\mathrm{c},i} w_{\mathrm{c},j}|\leq \bar v_i(h_{\mathrm{c},i})-|\tilde A_i\bar x+\tilde B_i\bar u|\] 
    for all $i\in\mathcal{I}(H_\mathrm{c})$, which implies that $\hat v_j\in\mathcal{V}(H_\mathrm{c})$.\\
Secondly we show that $\hat x_j\in\mathcal{E}(P_\mathrm{c}/\gamma_\mathrm{c})\oplus \bar x$ and $w_{\mathrm{c},j}\in\mathcal{E}(Q_{w_\mathrm{c}}^0)$ imply that $\hat x_j^+\in\mathcal{E}(P_\mathrm{c}/\gamma_\mathrm{c})\oplus\bar{x}$. To do it consider two system motions, i.e., $\hat x_j$, generated according to system \eqref{eq:closed_loop_observer} with noise $w_{\mathrm{c},j}$, and $\bar{x}$, generated according to system \eqref{eq:closed_loop_observer} with noise $\bar{w}_c=0$ and initial condition $\bar{x}$. Since, as shown above, both $\bar{v}\in\mathcal{V}(H_\mathrm{c})$ and ${\hat v}_j\in\mathcal{V}(H_\mathrm{c})$, as proved in step 1, $V(\hat x_j^+,\bar{x})-V(\hat x_j,\bar{x})=\|\hat x_j^+-\bar{x}\|^2_{P_\mathrm{c}}-\|\hat x_j-\bar{x}\|^2_{P_\mathrm{c}}\leq-\norm{\hat x_j-\bar{x}}^2_{Q_{\mathrm{c},x}}+\norm{w_{\mathrm{c},j}}^2_{Q_{c,w_\mathrm{c}}}$.
In view of Assumption~\ref{ass:disturbance}, condition \eqref{eq:mRPI_set_condition_1_dISS} implies $\norm{w_{\mathrm{c},j}}^2_{Q_{c,w_\mathrm{c}}}\leq \norm{w_{\mathrm{c},j}}^2_{Q_{w_\mathrm{c}}^0}\leq 1$. 
This implies that
\begin{equation}
V(\hat x_j^+,\bar{x})-V(\hat x_j,\bar{x})=\|\hat x_j^+\!-\!\bar{x}\|^2_{P_\mathrm{c}}-\|\hat x_j\!-\!\bar{x}\|^2_{P_\mathrm{c}}
\leq-\norm{\hat x_j-\bar{x}}^2_{Q_{\mathrm{c},x}}+1.
\label{eq:dissipation_form_function_4}
\end{equation}
Moreover, note that applying a congruence transformation to  \eqref{eq:mRPI_set_condition_2_dISS} yields
\(Q_{\mathrm{c},x}-P_\mathrm{c}/\gamma_\mathrm{c}\succeq0\).
Therefore, since \eqref{eq:dissipation_form_function_4} and \eqref{eq:mRPI_set_condition_2_dISS} hold, it follows from Lemma~\ref{lem:RPI_set} that if $\hat x-\bar x\in\mathcal{E}(P_\mathrm{c}/\gamma_\mathrm{c})$, then $\hat x^+-\bar x\in\mathcal{E}(P_\mathrm{c}/\gamma_\mathrm{c})$, i.e., $\hat x^+\in\mathcal{E}(P_\mathrm{c}/\gamma_\mathrm{c})\oplus\bar x$, as claimed.\smallskip\\
%
%
3. Note that, under the control law \eqref{eq:control_law_tracking}, $\tilde A\hat x+\tilde Bu=\tilde A_\mathrm{K}(\hat x-\bar x)+\tilde A\bar x+\tilde B\bar u$. Thus, to verify \eqref{eq:locality_condition_obs}, it is sufficient to show that, under \eqref{eq:locality_set_dISS_ctrobs1}, \eqref{eq:locality_set_dISS_ctrobs2}, and \eqref{eq:locality_eq_dISS_ctrobs}, the conditions $\tilde A_\mathrm{K}(\hat x-\bar x)+\tilde Ae+\tilde Dw\in\mathcal{V}(H_\mathrm{o})\ominus|\tilde A\bar x+\tilde B\bar u|$ and $\tilde A_\mathrm{K}(\hat x-\bar x)+\tilde LCe+\tilde L\eta\in\mathcal{V}(H_\mathrm{o})\ominus|\tilde A\bar x+\tilde B\bar u|$ hold.\\
First, note that \eqref{eq:locality_eq_dISS_ctrobs} implies that the set $\mathcal{V}(H_\mathrm{o})\ominus|\tilde A\bar x+\tilde B\bar u|$ is non-empty. 
Next, left- and right-multiplying condition \eqref{eq:locality_set_dISS_ctrobs1} by $\diag(P_\mathrm{c},I_{n+d+1})$, and recalling that $Z=KQ_\mathrm{c}$ and $\tilde A_\mathrm{K}=\tilde A+\tilde BK$, it follows that, for all $i\in\mathcal{I}(H_\mathrm{o})$,
\begin{equation*}
\begin{bmatrix} P_\mathrm{c}/(3\gamma_\mathrm{c})&0&0&\tilde A_{\mathrm{K},i}^\top\\
        0&P_\mathrm{o}/(3\gamma_\mathrm{o})&0&\tilde A_{i}^\top\\
        0&0&Q_w^0/3&\tilde D_{i}\\
        \tilde A_{\mathrm{K},i}&\tilde A_{i}&\tilde D_{i}&(\bar v_i(h_{\mathrm{o},i})-|\tilde A_i\bar x+\tilde B_i\bar u|)^2
        \end{bmatrix}\succeq 0.
\end{equation*}
By Lemma~\ref{lem:constr_satisfaction}, this last condition implies that for $(\hat x-\bar x,e,w)\in\mathcal{E}(P_\mathrm{c}/\gamma_\mathrm{c})\times\mathcal{E}(P_\mathrm{o}/\gamma_\mathrm{o})\times\mathcal{E}(Q_w^0)$, the inequality $|\tilde A_{\mathrm{K},i}(\hat x-\bar x)+\tilde A_ie+\tilde D_iw|\leq\bar{v}_i(h_{\mathrm{o},i})-|\tilde A_i\bar x+\tilde B_i\bar u|$ holds for all $i\in\mathcal{I}(H_\mathrm{o})$, which is equivalent to $\tilde A_\mathrm{K}(\hat x-\bar x)+\tilde Ae+\tilde Dw\in\mathcal{V}(H_\mathrm{o})\ominus|\tilde A\bar x+\tilde B\bar u|$.\\
Similarly, left- and right-multiplying condition \eqref{eq:locality_set_dISS_ctrobs2} by $\diag(P_\mathrm{c},I_{n+p+1})$, it follows that, for all $i\in\mathcal{I}(H_\mathrm{o})$,
\begin{equation*}
\begin{bmatrix}\ P_\mathrm{c}/(3\gamma_\mathrm{c})\!\!&\!\!0&\!\!0\!\!&\!\!\tilde A_{\mathrm{K},i}^\top\\
        0\!\!&\!\!P_\mathrm{o}/(3\gamma_\mathrm{o})\!\!&\!\!0\!\!&\!\!C^\top\tilde L_{i}^\top\\
        0\!\!&\!\!0\!\!&\!\!Q_\eta^0/3\!\!&\!\!\tilde L_{i}^\top\\
        \tilde A_{\mathrm{K},i}\!\!&\!\!\tilde L_{i}C\!\!&\!\!\tilde L_{i}\!\!&\!\!(\bar v_i(h_{\mathrm{o},i})\!-\!|\tilde A_i\bar x\!+\!\tilde B_i\bar u|)^2
        \end{bmatrix}\!\!\succeq 0,
\end{equation*}
which, by Lemma~\ref{lem:constr_satisfaction}, implies that for $(\hat x-\bar x,e,\eta)\in\mathcal{E}(P_\mathrm{c}/\gamma_\mathrm{c})\times\mathcal{E}(P_\mathrm{o}/\gamma_\mathrm{o})\times\mathcal{E}(Q_\eta^0)$, the inequality $|\tilde A_{\mathrm{K},i}(\hat x-\bar x)+\tilde L_iCe+\tilde L_in|\leq\bar{v}_i(h_{\mathrm{o},i})-|\tilde A_i\bar x+\tilde B_i\bar u|$ holds for all $i\in\mathcal{I}(H_\mathrm{o})$, proving that $\tilde A_\mathrm{K}(\hat x-\bar x)+\tilde LCe+\tilde L\eta\in\mathcal{V}(H_\mathrm{o})\ominus|\tilde A\bar x+\tilde B\bar u|$.

4. Now, consider two generic motions $\hat x_1$ and $\hat x_2$. In case $\mathcal I(H_\mathrm{c}) =\emptyset$, the Lyapunov function $V(\hat x_1,\hat x_2)$ satisfies \eqref{eq:dissipation_form_function} for all $\hat x_1, \hat x_2\in\R^n$ and $w_{\mathrm{c},1}, w_{\mathrm{c},2}\in\mathcal{E}(Q_{w_\mathrm{c}}^0)$. Additionally, note that $\lambda_{\min}(P_\mathrm{c})\norm{\hat x_1-\hat x_2}^2\leq V(\hat x_1,\hat x_2)\leq\lambda_{\max}(P_\mathrm{c})\norm{\hat x_1-\hat x_2}^2$. Also, $\alpha_1(\norm{\hat x_1-\hat x_2})=\lambda_{\min}(P_\mathrm{c})\norm{\hat x_1-\hat x_2}^2$, $\alpha_2(\norm{\hat x_1-\hat x_2})=\lambda_{\max}(P_\mathrm{c})\norm{\hat x_1-\hat x_2}^2$ and $\alpha_3(\norm{\hat x_1-\hat x_2})=\norm{\hat x_1-\hat x_2}^2_{Q_{\mathrm{c},x}}$ are class $\mathcal{K}_\infty$-functions, and $\alpha_4(\norm{w_{\mathrm{c},1}-w_{\mathrm{c},2}})=\norm{w_{\mathrm{c},1}-w_{\mathrm{c},2}}^2_{Q_{c,w_\mathrm{c}}}$ is a class $\mathcal{K}$-function. Therefore, $V(\hat x_1,\hat x_2)$ is a dissipation-form $\delta$ISS function over the sets $\R^n$ and $\mathcal{E}(Q_{w_\mathrm{c}}^0)$. According to Theorem~\ref{th:local_dISS}, system \eqref{eq:closed_loop_observer} is $\delta$ISS with respect to these sets.\\
   On the other hand, in case $\mathcal I(H_\mathrm{c}) \neq\emptyset$, due to the invariance of the set $\mathcal{E}(P_\mathrm{c}/\gamma_\mathrm{c})\oplus\bar x$, if the initial conditions satisfy $\hat x_1(0),\hat x_2(0)\in\mathcal{E}(P_\mathrm{c}/\gamma_\mathrm{c})\oplus\bar x$, then for all $k>0$ we have $\hat x_1(k)\in\mathcal{E}(P_\mathrm{c}/\gamma_\mathrm{c})\oplus\bar x$ and $\hat x_2(k)\in\mathcal{E}(P_\mathrm{c}/\gamma_\mathrm{c})\oplus\bar x$. Since $\mathcal{E}(P_\mathrm{c}/\gamma_\mathrm{c})\oplus\bar x\subseteq\mathcal{V}(H_\mathrm{c})$ and, by assumption, $w_{\mathrm{c},1},w_{\mathrm{c},2}\in\mathcal{E}(Q_{w_\mathrm{c}}^0)$, it follows that $\hat v_1,\hat v_2\in\mathcal V(H_\mathrm{c})$. Thus, $V(\hat x_1,\hat x_2)$ satisfies \eqref{eq:dissipation_form_function}, which, by Theorem~\ref{th:local_dISS}, implies that system \eqref{eq:closed_loop_observer} is $\delta$ISS with respect to the sets $\mathcal{E}(P_\mathrm{c}/\gamma_\mathrm{c})\oplus\bar x$ and $\mathcal{E}(Q_{w_\mathrm{c}}^0)$. This concludes the proof.\hfill{}$\square$\smallskip\\
\textbf{Proof of Lemma \ref{lem:NScond}.}
We first prove that the feasibility of \eqref{eq:dISS_condition} implies the stabilizability of $(A,B)$.
Recalling that $Z = KQ_\mathrm{c}$, \eqref{eq:dISS_condition} is equivalent to \eqref{eq:dISS_condition_congr} which, applying a congruence transformation with \eqref{eq:congr_transf}, yields
    \begin{equation}\label{eq:dISS_condition_2}
    \begin{bmatrix}
    Q_{\mathrm{c}}-\tilde Q_{\mathrm{c},x}&Q_\mathrm{c}A_\mathrm{K}^\top&-Q_{\mathrm{c}}\tilde A_\mathrm{K}^\top&0\\
    A_\mathrm{K}Q_\mathrm{c}&Q_{\mathrm{c}}&B_qU_\mathrm{c}&D_{\mathrm{c}}\\
    -\tilde A_\mathrm{K}Q_\mathrm{c}&U_\mathrm{c}B_q^\top&2H_\mathrm{c}U_\mathrm{c}&-\tilde D_{\mathrm{c}}\\
    0&D_{\mathrm{c}}^\top&-\tilde D_{\mathrm{c}}^\top&Q_{c,w_\mathrm{c}}
\end{bmatrix}\succeq 0.
\end{equation}    
Note that a necessary condition for \eqref{eq:dISS_condition_2} is that
\begin{equation}\label{eq:necessary_condition_1}
\begin{bmatrix}
    Q_{\mathrm{c}}-\tilde Q_{\mathrm{c},x}&Q_\mathrm{c}A_\mathrm{K}^\top\\
    A_\mathrm{K}Q_\mathrm{c}&Q_{\mathrm{c}}
\end{bmatrix}\succeq 0.
\end{equation}
By left- and right-multiplying condition \eqref{eq:necessary_condition_1} by $\diag(P_\mathrm{c},P_\mathrm{c})$, and recalling that $P_\mathrm{c}=Q_{\mathrm{c}}^{-1}$ and $Q_{\mathrm{c},x}=P_\mathrm{c}\tilde Q_{\mathrm{c},x}P_\mathrm{c}$, we obtain 
\begin{equation}\label{eq:necessary_condition_2}
\begin{bmatrix}
    P_\mathrm{c}-Q_{\mathrm{c},x}&A_\mathrm{K}^\top P_\mathrm{c}\\
    P_\mathrm{c}A_\mathrm{K}&P_\mathrm{c}
\end{bmatrix}\succeq 0.
\end{equation}
Since $P_\mathrm{c}$ is invertible by design, we can apply the Schur complement to  \eqref{eq:necessary_condition_2}, yielding the condition $P_\mathrm{c}-Q_{\mathrm{c},x}-A_\mathrm{K}^\top P_\mathrm{c}A_\mathrm{K}\succeq 0$. This inequality holds only if $A_\mathrm{K}$ is a Schur matrix, having quadratic Lyapunov function $V(x)=x^\top P_\mathrm{c}x$, with $x\in\R^n$. This is possible only if $(A,B)$ is stabilizable.\smallskip\\
We now show that the stabilizability of $(A,B)$ implies the feasibility of \eqref{eq:dISS_condition}. By applying the Schur complement, condition \eqref{eq:dISS_condition_2} can be rewritten as
\begin{equation}\label{eq:dISS_condition_3}
    \begin{bmatrix}
    Q_{\mathrm{c}}-\tilde Q_{\mathrm{c},x}&Q_\mathrm{c}A_\mathrm{K}^\top\\
    A_\mathrm{K}Q_\mathrm{c}&Q_{\mathrm{c}}
\end{bmatrix}-M_1 Q_{Hw_\mathrm{c}}^{-1}M_1^\top\succeq0
\end{equation}
where
\begin{equation*}
M_1=\begin{bmatrix}
    -Q_{\mathrm{c}}\tilde A_\mathrm{K}^\top&0\\
    B_qU_\mathrm{c}&D_{\mathrm{c}}
\end{bmatrix}, \ \ Q_{Hw_\mathrm{c}}=\begin{bmatrix}
    2H_\mathrm{c}U_\mathrm{c}&-\tilde D_{\mathrm{c}}\\
    -\tilde D_{\mathrm{c}}^\top&Q_{c,w_\mathrm{c}}
\end{bmatrix}.
\end{equation*}
Note that it implies the existence of $K$ such that $A_\mathrm{K}$ is Schur stable.
The rest of the proof follows similar steps to the proof of Lemma \ref{lem:NScond_obs}. In particular, a sufficient condition for \eqref{eq:dISS_condition_3} can be derived by selecting $Q_{\mathrm{c}}\succ0$ such that
\[\begin{bmatrix}
    Q_{\mathrm{c}}-\tilde Q_{\mathrm{c},x}&Q_\mathrm{c}A_\mathrm{K}^\top\\
    A_\mathrm{K}Q_\mathrm{c} &Q_{\mathrm{c}} 
\end{bmatrix}\succ0,\]
which requires the Schur stability of $A_\mathrm{K}$,
and, for instance, $U_\mathrm{c} = \frac{1}{2} I_\nu$, $Q_{c,w_\mathrm{c}} = \mu I_d$,  $H_\mathrm{c} = \mu I_\nu$, and a sufficiently large $\mu>0$.
\hfill{}$\square$\smallskip\\
\textbf{Proof of Theorem~\ref{th:closed_loop_K_as_gain}.}
To prove Theorem~\ref{th:closed_loop_K_as_gain}, note that under Assumptions~\ref{ass:disturbance} and \ref{ass:output_noise}, and assuming $e(0) \in \mathcal{E}(P_\mathrm o / \gamma_o)$, Proposition~\ref{prop:dISS_observer} implies the existence of a class $\mathcal{KL}$-function $\beta_\mathrm o$ and class $\mathcal{K}$-functions $\gamma_{\mathrm o,w}$ and $\gamma_{\mathrm o,\eta}$ such that system~\eqref{eq:observer_dynamics} satisfies
\begin{equation*}
    \norm{e(k)}\leq\beta_o(\norm{e(0)},k)        +\gamma_{\mathrm o,w}(\max_{h\geq 0}\norm{w(h)})+\gamma_{\mathrm o,\eta}(\max_{\eta\geq 0}\norm{\eta(h)}).
\end{equation*}
Following similar steps as in the proof of Lemma $3.8$ in \cite{jiang2001input}, it is possible to show that this condition implies the existence of a class $\mathcal{K}$-function $\gamma_{\mathrm o}$ such that
\begin{equation}\label{eq:k_as_gain_obs}
    \overline{\lim}_{k \to+ \infty} \norm{e(k)}\leq\gamma_{\mathrm o}(\overline{\lim}_{k \to+ \infty} \norm{w(k)})+\gamma_{\mathrm o}(\overline{\lim}_{k \to+ \infty}\norm{\eta(k)}).
\end{equation}
Similarly, Proposition~\ref{prop:dISS} guarantees the existence of functions $\beta_c\in\mathcal{KL}$ and $\gamma_{c,\eta},\gamma_{c,e}\in\mathcal{K}_\infty$, such that system~\eqref{eq:closed_loop_observer} satisfies
\begin{equation*}
    \norm{\hat x(k)-\bar x}\leq\beta_\mathrm c(\norm{\hat x(k)-\bar x},k)        +\gamma_{\mathrm c,\eta}(\max_{h\geq 0}\norm{\eta(h)})+\gamma_{\mathrm c,e}(\max_{h\geq 0}\norm{e(h)}),
\end{equation*}
which implies the existence of a class $\mathcal{K}$-function $\gamma_{\mathrm c}$ such that
\begin{equation}\label{eq:k_as_gain_control}
    \overline{\lim}_{k \to+ \infty} \norm{\hat x(k)-\bar x}\leq\gamma_{\mathrm c}(\overline{\lim}_{k \to+ \infty} \norm{\eta(k)})+\gamma_{\mathrm c}(\overline{\lim}_{k \to+ \infty}\norm{e(k)}).
\end{equation}
Noting that $\norm{x-\bar x}=\norm{\hat x-\bar x+e}\leq\norm{\hat x-\bar x}+\norm{e}$, and
using \eqref{eq:k_as_gain_obs} and \eqref{eq:k_as_gain_control}, yields
\begin{equation*}
    \begin{aligned}
        \overline{\lim}_{k \to+ \infty} \norm{x(k)-\bar x}
        &\leq\overline{\lim}_{k \to+ \infty} \norm{\hat x(k)-\bar x}+\overline{\lim}_{k \to+ \infty} \norm{e(k)}\\
        &\leq\gamma_{\mathrm c}(\overline{\lim}_{k \to+ \infty} \norm{\eta(k)})+\gamma_{c}(\overline{\lim}_{k \to+ \infty}\norm{e(k)})
        +\overline{\lim}_{k \to+ \infty} \norm{e(k)}\\
        &\leq\gamma_{\mathrm c}(\overline{\lim}_{k \to+ \infty} \norm{\eta(k)})+\gamma_{\mathrm c}(\gamma_{\mathrm o}(\overline{\lim}_{k \to+ \infty} \norm{w(k)})+\gamma_{\mathrm o}(\overline{\lim}_{k \to+ \infty}\norm{\eta(k)}))\\&\quad  +\gamma_{\mathrm o}(\overline{\lim}_{k \to+ \infty} \norm{w(k)})+\gamma_{\mathrm o}(\overline{\lim}_{k \to+ \infty}\norm{\eta(k)})\\
        &\leq \gamma_{\mathrm c}(\overline{\lim}_{k \to+ \infty} \norm{\eta(k)})+\gamma_{\mathrm{co}}(\overline{\lim}_{k \to+ \infty} \norm{w(k)})+\gamma_{\mathrm{co}}(\overline{\lim}_{k \to+ \infty}\norm{\eta(k)})\\&\quad+\gamma_{\mathrm o}(\overline{\lim}_{k \to+ \infty}\norm{w(k)}) +\gamma_{\mathrm o}(\overline{\lim}_{k \to+ \infty} \norm{\eta(k)})).
    \end{aligned}
\end{equation*}
where, according to the weak triangle inequality~\cite{kellett2014compendium}, $\gamma_{\mathrm {co}}(\cdot)=\gamma_\mathrm c(2\gamma_\mathrm o(\cdot))\in\mathcal{K}$.
Finally, by leveraging the super-additivity property of class $\mathcal{K}$-functions \cite{wiltz2024note}, it can be shown that there exist functions $\gamma_w\in \mathcal{K}$ such that 
\[\gamma_{\mathrm{co}}(\overline{\lim}_{k \to + \infty} \|w(k)\|)+\gamma_{o}(\overline{\lim}_{k \to + \infty} \|w(k)\|) \leq \gamma_w(\overline{\lim}_{k \to + \infty} \|w(k)\|),\]
and function $\gamma_\eta \in \mathcal{K}$ such that
\[\gamma_{\mathrm c}(\overline{\lim}_{k \to + \infty} \|\eta(k)\|) + \gamma_{\mathrm{co}}(\overline{\lim}_{k \to + \infty} \|\eta(k)\|)) + \gamma_{\mathrm o}(\overline{\lim}_{k \to + \infty} \|\eta(k)\|) \leq \gamma_\eta(\overline{\lim}_{k \to + \infty} \|\eta(k)\|),\]
therefore proving \eqref{eq:closed_loop_K_as_gain}.
\hfill{}$\square$\smallskip\\
\textbf{Proof of Proposition \ref{prop:dISS_control}.}
The first part of the proof of Proposition~\ref{prop:dISS_control} follows from Proposition~\ref{prop:dISS}.\\
Note that the dynamics of the closed-loop observer \eqref{eq:closed_loop_observer_MPC} is analogous to \eqref{eq:closed_loop_observer}. Therefore, under Assumptions \ref{ass:sigmoid_function}, \ref{ass:disturbance} and \ref{ass:output_noise}, and for $e(k)\in\mathcal{E}(P_\mathrm{o}/\gamma_\mathrm{o})$, Proposition~\ref{prop:dISS} guarantees that if conditions \eqref{eq:dISS_condition}-\eqref{eq:mRPI_set_condition_2_dISS} hold with $\mathcal I(H_\mathrm{c})=\emptyset$, then system \eqref{eq:closed_loop_observer_MPC} is $\delta$ISS with respect to the sets $\R^n$ and $\mathcal{E}(Q_{w_\mathrm{c}}^0)$. On the other hand, if conditions \eqref{eq:dISS_condition}-\eqref{eq:mRPI_set_condition_2_dISS} hold with $\mathcal I(H_\mathrm{c})\neq\emptyset$ and if, for $\hat x\in\mathcal{E}(P_\mathrm{c}/\gamma_\mathrm{c})\oplus \tilde x$, the locality constraints are satisfied, i.e. $\hat v,\tilde A\tilde x+ \tilde B\tilde u\in\mathcal V(H_\mathrm{c})$, then system \eqref{eq:closed_loop_observer_MPC} is $\delta$ISS with respect to the sets $\mathcal{E}(P_\mathrm{c}/\gamma_\mathrm{c})\oplus \tilde x$ and $\mathcal{E}(Q_{w_\mathrm{c}}^0)$.  Also, the set $\mathcal{E}(P_\mathrm{c}/\gamma_\mathrm{c})\oplus \tilde x$ is RPI for the system dynamics.\\
Now, we focus on the case where $\mathcal{I}(H_\mathrm{c})\neq\emptyset$. Noting that $\hat v=\tilde A_\mathrm{K}\hat x+\tilde B(\tilde u-K\tilde x)+\tilde D_{\mathrm{c}}w_\mathrm{c}=\tilde A\tilde  x+\tilde B\tilde u+\tilde A_\mathrm{K}(\hat x-\tilde x)+\tilde D_{\mathrm{c}}w_\mathrm{c}$, condition \eqref{eq:locality_constr_MPC} is sufficient to ensure that $\hat v,\tilde A\tilde x+\tilde B\tilde u\in\mathcal V(H_\mathrm{c})$. 
The set $\bar{\mathcal V}_{\mathrm{c}}$ exists only if $(\tilde e,w_\mathrm{c})\in \mathcal E(P_\mathrm{c}/\gamma_\mathrm{c})\times\mathcal E(Q_{w_\mathrm{c}}^0)$ implies $\tilde A_\mathrm{K}\tilde e\oplus\tilde D_{\mathrm{c}}w_\mathrm{c}\in\mathcal{V}(H_\mathrm{c})$, i.e.
$|\tilde A_\mathrm{K}\tilde e+\tilde D_{\mathrm{c}}w_\mathrm{c}|\leq\bar v_i(h_{\mathrm{c},i})$, $\forall i\in\mathcal{I}(H_\mathrm{c})$,
which, by Lemma~\ref{lem:constr_satisfaction}, holds if
\begin{equation}\label{eq:locality_set_dISS_mpc_2}
\begin{bmatrix}\
        P_\mathrm{c}/(2\gamma_\mathrm{c})&0&\tilde A_{\mathrm{K},i}^\top\\
        0&Q_{w_\mathrm{c}}^0/2&\tilde D_{\mathrm{c},i}\\
        \tilde A_{\mathrm{K},i}&\tilde D_{\mathrm{c},i}&\bar v_i(h_{\mathrm{c},i})^2
        \end{bmatrix}\succeq 0, \ \forall i\in\mathcal{I}(H_\mathrm{c}).
\end{equation}
Finally, left- and right-multiplying \eqref{eq:locality_set_dISS_mpc_2} by  $M_\mathrm f = \diag{(Q_{\mathrm{c}},I_{n+p+1})}$, yields \eqref{eq:locality_set_dISS_mpc}.
\hfill{}$\square$\smallskip\\
\textbf{Proof of Lemma \ref{lem:MPC_locality_cstr}.}
Consider the problem of ensuring \eqref{eq:locality_condition_obs} when $\tilde e\in\mathcal{E}(P_\mathrm{c}/\gamma_\mathrm{c})$. 
Using \eqref{eq:control_law_tube} and noting that $\hat{x} = \tilde{e} + \tilde{x}$, it follows that $\tilde{A} \hat{x} + \tilde{B} u = \tilde{A} \tilde{x} + \tilde{B} \tilde{u} + \tilde{A}_K \tilde{e}$. Consequently, \eqref{eq:locality_constr_obsMPC} provides a sufficient condition for \eqref{eq:locality_condition_obs} to hold.
A necessary condition for the set $\bar{\mathcal{V}}_{o_c}$ to exist is that $(e,\tilde e,w,\eta)\in\mathcal E(P_\mathrm{o}/\gamma_\mathrm{o})\times\mathcal E(P_\mathrm{c}/\gamma_\mathrm{c})\times\mathcal E(Q_w^0)\times\mathcal E(Q_\eta^0)$ implies $\tilde Ae+\tilde Dw+\tilde A_\mathrm{K}\tilde e\in\mathcal V(H_\mathrm{o})$ and $\tilde LCe+\tilde L\eta+\tilde A_\mathrm{K}\tilde e\in\mathcal V(H_\mathrm{o})$.\\
According to Lemma~\ref{lem:constr_satisfaction}, the first implication holds if, for all $i\in\mathcal I(H_\mathrm{o})$, 
\begin{equation*}
    \begin{bmatrix}
    P_\mathrm{o}/(3\gamma_\mathrm{o})&0&0&\tilde A_i^\top\\
    0&P_\mathrm{c}/(3\gamma_\mathrm{c})&0&\tilde A_{\mathrm{K},i}^\top&\\
    0&0&Q_w^0/3&\tilde D_i^\top\\
    \tilde A_i&\tilde A_{\mathrm{K},i}&\tilde D_i&\bar v(h_{\mathrm{o},i})^2
    \end{bmatrix}\succeq 0,
\end{equation*}
from which, by applying a congruence transformation and noting that $\tilde A_{\mathrm{K},i}=\tilde A_i+\tilde B_iK$ and $Z=KP_\mathrm{c}^{-1}$, condition \eqref{eq:locality_set_dISS_obs_mpc_1} is derived.\\
Using similar arguments, it follows that the second implication holds if, for all $i\in\mathcal I(H_\mathrm{o})$, 
\begin{equation*}
    \begin{bmatrix}
    P_\mathrm{o}/(3\gamma_\mathrm{o})&0&0&C^\top\tilde L_i^\top\\
    0&P_\mathrm{c}/(3\gamma_\mathrm{c})&0&\tilde A_{\mathrm{K},i}^\top&\\
    0&0&Q_\eta^0/3&\tilde L_i^\top\\
    \tilde L_iC&\tilde A_{\mathrm{K},i}&\tilde L_i&\bar v(h_{\mathrm{o},i})^2
    \end{bmatrix}\succeq 0,
\end{equation*}
which, by applying a congruence transformation, is equivalent to \eqref{eq:locality_set_dISS_obs_mpc_2}.
\hfill{}$\square$\smallskip\\
\textbf{Proof of Theorem~\ref{th:FHOCP_guarantees}.}
The proof of Theorem~\ref{th:FHOCP_guarantees} uses standard arguments. Specifically, we verify that the three main assumptions of \cite[Theorem 3]{bayer2013discrete} are fulfilled:
\begin{itemize}
    \item A1: $C\mathbb{X}_\mathrm{f}(\bar x)\in\tilde{\mathbb{Y}}$, $\mathbb{X}_\mathrm{f}(\bar x)$ is closed, $\bar{x}\in\mathbb{X}_\mathrm{f}(\bar x)$;
    \item A2: $\exists \kappa_\mathrm{f}(\tilde{x})$ such that $\kappa_\mathrm{f}(\tilde{x})\in \tilde{\mathbb U}$ and $f(A\hat{x}+B\kappa_\mathrm{f}(\hat{x}))\in \mathbb{X}_\mathrm{f}(\bar x)$ $\forall \tilde{x}\in \mathbb{X}_\mathrm{f}(\bar x)$;
    \item A3: $V_\mathrm{f}(f(A\tilde{x}+B\kappa_\mathrm{f}(\tilde{x})))-V_\mathrm{f}(\tilde{x})\leq -(\|\tilde{x}-\bar{x}\|^2_{\Lambda_x}+\|\kappa_\mathrm{f}(\tilde{x})-\bar{u}\|^2_{\Lambda_u})$,  $\forall \tilde x\in\mathbb{X}_\mathrm{f}(\bar x)$.
\end{itemize}
Moreover, in case the conditions of Proposition~\ref{prop:dISS_observer} or Proposition~\ref{prop:dISS_control} hold locally, meaning  $\mathcal I(H_\mathrm{o})\neq\emptyset$ or $\mathcal I(H_\mathrm{c})\neq\emptyset$, respectively, it is also necessary to verify the following assumption. \\A4: For all $\tilde x\in\mathbb{X}_\mathrm{f}(\bar x)$, it holds that $\tilde A\tilde x+\tilde B\kappa_\mathrm{f}(\tilde{x})\in\mathcal{V}_\mathrm{o}(H_\mathrm{o})\cap\mathcal{V}_{\mathrm{c}}(H_\mathrm{c})$.
\smallskip\\
To prove the first claim of A1, assume that $\tilde x\in\mathbb{X}_\mathrm{f}(\bar x)$, i.e. $\tilde x-\bar x\in\mathcal{E}(P_\mathrm{f}/\gamma_\mathrm{f})$. By Lemma~\ref{lem:constr_satisfaction} and recalling that $\tilde\gamma_\mathrm{f}=1/\gamma_\mathrm{f}$, condition \eqref{eq:output_constr_terminal} implies $|G_{\mathrm{y},r}C(\tilde x-\bar x)|\leq b_{\mathrm{y},r}-G_{\mathrm{y},r}C\bar x$, for all $r\in\{1,\dots,n_r\}$, which further implies $G_{\mathrm{y},r}C(\tilde x-\bar x)\leq b_{\mathrm{y},r}-G_{\mathrm{y},r}C\bar x$, for all $r\in\{1,\dots,n_r\}$. Rearranging the last inequality, we obtain $G_{\mathrm{y},r}C\tilde x\leq b_{\mathrm{y},r}$, for all $r\in\{1,\dots,n_r\}$, which is equivalent to $C\tilde x\in\tilde {\mathbb{Y}}$. Moreover, the ellipsoidal set $\mathbb{X}_\mathrm{f}(\bar x)$ is closed, as $P_\mathrm{f}\succ0$, and, by definition, it contains $\bar x$.\\
Similarly, condition \eqref{eq:input_constr_terminal} implies $|G_{\mathrm{u},s}K(\tilde x-\bar x)|\leq b_{\mathrm{u},s}-G_{\mathrm{u},s}\bar u$, for all $s\in\{1,\dots,n_s\}$, which further implies $G_{\mathrm{u},s}K(\tilde x-\bar x)\leq b_{\mathrm{u},s}-G_{\mathrm{u},s}\bar u$, for all $s\in\{1,\dots,n_s\}$. Rearranging the last inequality and using \eqref{eq:control_law_aux}, we obtain $G_{\mathrm{u},s}\tilde u\leq b_{\mathrm{u},s}$, for all $s\in\{1,\dots,n_s\}$, which satisfies the first claim of Assumption A2.\\
To prove A3, note that the nominal dynamics \eqref{eq:nom_system} under the auxiliary control law \eqref{eq:control_law_aux} is
    $\tilde x(k+1)=A_{K}\tilde x(k)+B(\bar u-K\bar x)+B_qq(\tilde A_{\mathrm{K}}\tilde x(k)+\tilde B(\bar u-K\bar x))$,
which is equivalent to \eqref{eq:closed_loop_observer} in case of zero disturbance. 
Therefore, considering the Lyapunov function $V_\mathrm{f}(\tilde x(k),\bar x)=\norm{\tilde x(k)-\bar x}_{P_\mathrm{f}}^2$, condition \eqref{eq:dISS_terminal} can be derived from \eqref{eq:dISS_condition} in case of zero disturbance and setting $Q_{\mathrm{c},x}=\Lambda_x+K^T\Lambda_uK$. According to the proof of Proposition~\ref{prop:dISS}, it follows that \eqref{eq:dISS_terminal} implies that, for all $\tilde x$ such that $A_{K}\tilde x\in \mathcal V(H_\mathrm{f})\ominus|\tilde A\bar{x}+\tilde B\bar{u}|$,
\begin{equation}\label{eq:NMPC_ly_function}
V_\mathrm{f}(\tilde x(k+1),\bar x)-V_\mathrm{f}(\tilde x(k),\bar x)\leq-\norm{\tilde x(k)-\bar x}_{\Lambda_x+K^\top\Lambda_uK}^2.
\end{equation}
If $\mathcal I(H_\mathrm{f})=\emptyset$, then $\mathcal V(H_\mathrm{f})=\R^\nu$. On the other hand, if $\mathcal I(H_\mathrm{f})\neq\emptyset$, then, according to Lemma~\ref{lem:constr_satisfaction}, condition \eqref{eq:locality_constr_hf_terminal} implies that for all $\tilde x-\bar x\in\mathcal{E}(P_\mathrm{f}/\gamma_\mathrm{f})$, it holds that $\tilde A_{\mathrm{K},i}\tilde x\in\bar v_i(h_{\mathrm{f},i})-|\tilde A_i\bar{x}+\tilde B_i\bar{u}|$, for all $i\in\mathcal I(H_\mathrm{f})$, i.e. $\mathcal{E}(P_\mathrm{f}/\gamma_\mathrm{f})\subseteq\mathcal V(H_\mathrm{f})\ominus|\tilde A\bar{x}+\tilde B\bar{u}|$.
Observing that $\norm{\tilde x(k)-\bar x}_{\Lambda_x+K^\top\Lambda_uK}^2=\norm{\tilde x(k)-\bar x}^2_{\Lambda_x}+\norm{K(\tilde x(k)-\bar x)}_{\Lambda_u}^2$, and noting that $K(\tilde x(k)-\bar x)=\tilde u(k)-\bar u$, it follows that A3 is satisfied.\\ Moreover, since \eqref{eq:NMPC_ly_function} implies $V_\mathrm{f}(\tilde x(k+1),\bar x)\leq V_\mathrm{f}(\tilde x(k),\bar x)$, the set $\mathcal E(P_\mathrm{f}/\gamma_\mathrm{f})$ is a forward invariant set for $\tilde x(k)-\bar x$, thereby satisfying the second claim of A2.\smallskip\\
%
Finally, we can prove A4 by showing that $\tilde x-\bar x\in\mathcal{E}(P_\mathrm{f}/\gamma_\mathrm{f})$ implies $\tilde A\tilde x+\tilde B\tilde u\in \mathcal{V}_{\mathrm{c}}(H_\mathrm{c})\ominus|\tilde A\bar x+\tilde B\bar u|$ and $\tilde A\tilde x+\tilde B\tilde u\in \mathcal{V}_\mathrm{o}(H_\mathrm{o})\ominus|\tilde A\bar x+\tilde B\bar u|$. By Lemma~\eqref{lem:constr_satisfaction}, condition \eqref{eq:locality_constr_hc_terminal} (resp., \eqref{eq:locality_constr_ho_terminal}) implies $|\tilde A_i\tilde x+\tilde B_i\tilde u|\leq\bar v_{\mathrm{c},i}(h_\mathrm{c},i)-|\tilde A_i\bar{x}+\tilde B_i\bar{u}|$, for all $i\in\mathcal{I}(H_\mathrm{c})$, (resp., $|\tilde A_i\tilde x+\tilde B_i\tilde u|\leq\bar v_{\mathrm{o},i}^\star(h_{\mathrm{o},i})-|\tilde A_i\bar{x}+\tilde B_i\bar{u}|$, for all $i\in\mathcal{I}(H_\mathrm{o})$), completing the proof.
\hfill{}$\square$
%
%
\section{Appendix: Definitions of $\delta$ISS dynamics and ISS observer error dynamics}\label{appendix_B} %
Consider a generic nonlinear discrete-time system described by
\begin{equation}\label{eq:nonlinear_sys}
\begin{cases}
    x(k+1)=f(x(k),w(k))\\
    y(k)=g({x}(k),\eta(k)).
\end{cases}
\end{equation}
where $k\in\mathbb Z_{\geq0}$ is the discrete-time index, $x\in\R^n$ is the state, $w\in\R^m$ is a perturbation acting on the system, $\eta\in \mathcal{N}\subseteq \mathbb{R}^p$ is the measurement noise, function  $f:\R^n\times\R^m\to\R^n$, and $g:\R^n\times\R^m\to\R^p$.
\begin{definition}[$\delta$ISS, \cite{bayer2013discrete}]\label{def:delta_ISS}
    System~\eqref{eq:nonlinear_sys} is said to be $\delta$ISS with respect to the sets $\mathcal{X}\subseteq\mathbb{R}^n$ and $\mathcal{W}\subseteq\mathbb{R}^m$ if $\mathcal{X}$ is robust positively invariant for \eqref{eq:nonlinear_sys} and there exist functions $\beta\in\mathcal{KL}$ and $\gamma\in\mathcal{K_\infty}$ such that for any $k\in\mathbb{Z}_{\geq0}$, any pair of initial states $x_a(0)\in\mathcal{X}$,  $x_b(0)\in\mathcal{X}$ and any pair of input $w_a\in\mathcal{W}$ and $w_b\in\mathcal{W}$, it holds that
    \begin{equation*}
    \norm{x_a(k)-x_b(k)}\leq\beta(\norm{x_a(0)-x_b(0)},k)+\gamma(\max_{h\geq 0}\norm{w_a(h)-w_b(h)}).
    \end{equation*}
    \hfill{}$\square$
\end{definition}
The $\delta$ISS property is a fundamental stability property, which can be guaranteed provided that a suitable $\delta$ISS Lyapunov function exists, as stated in the following \cite{bayer2013discrete}.
\begin{definition}[$\delta$ISS Lyapunov function]
A function $V(x_a,x_b)$ is a dissipation-form $\delta$ISS Lyapunov function over the sets $\mathcal X$ and $\mathcal W$ for system \eqref{eq:nonlinear_sys} if $\mathcal X$ is robust positively invariant for \eqref{eq:nonlinear_sys} and there exist $\mathcal{K_\infty}$-functions $\alpha_1$, $\alpha_2$ and $\alpha_3$, and a $\mathcal{K}$-function $\alpha_4$ such that, for any pair of states $x_a(k)\in\mathcal{X}$ and $x_b(k)\in\mathcal{X}$, and any pair of inputs $w_a(k)\in\mathcal{W}$ and $w_b(k)\in\mathcal{W}$, it holds that
    \begin{equation*}
       \alpha_1(\norm{x_a(k)-x_b(k)})\!\leq\! V(x_a(k),x_b(k))\!\leq\!\alpha_2(\norm{x_a(k)-x_b(k)}),
    \end{equation*}
    \begin{equation*}
    V(x_a(k+1),x_b(k+1))-V(x_a(k),x_b(k))\leq -\alpha_3(\norm{x_a(k)-x_b(k)})+\alpha_4(\norm{w_a(k)-w_b(k)}).
    \end{equation*}\hfill{}$\square$
\end{definition}
\begin{theorem}[\cite{bayer2013discrete}]\label{th:local_dISS}
   If system \eqref{eq:nonlinear_sys} admits a dissipation-form
    $\delta$ISS Lyapunov function over the sets $\mathcal{X}$ and $\mathcal{W}$, then it is $\delta$ISS with respect to such sets, in the sense of Definition~\ref{def:delta_ISS}.
    \hfill{}$\square$\end{theorem}
%
Suppose that an observer
\begin{equation}\label{eq:nonlinear_observer}
    \hat{x}(k+1)=f_\mathrm{o}(\hat{x}(k),y(k))
\end{equation}
for system \eqref{eq:nonlinear_sys} is available.
Define the estimation error as $e(k) \coloneq x(k)-\hat x(k)$. We introduce the following definition. %
\begin{definition}[ISS observer error dynamics]\label{def:delta_ISS_observer}
    Observer \eqref{eq:nonlinear_observer} is said to have ISS observer error dynamics with respect to the sets $\mathcal{X}_e\subseteq\R^n$, $\mathcal{W}\subseteq\R^m$ and $\mathcal{N}\subseteq \mathbb{R}^p$ if $\mathcal{X}_e$ is robust positively invariant for e(k) and if there exist functions $\beta_{\mathrm{o}}\in\mathcal{KL}$, $\gamma^w_\mathrm{o}\in\mathcal{K}$ and $\gamma^\eta_o\in\mathcal{K}$ such that for any $k\in\mathbb{Z}_{\geq0}$, any initial estimation error $e(0)\in\mathcal{X}_e$ and any pair of input and noise $w(k)\in\mathcal{W}$ and $\eta(k)\in\mathcal{N}$, it holds that
    \begin{equation*}
    \norm{e(k)}\leq\beta_{\mathrm{o}}(\norm{e(0)},k)        +\gamma_\mathrm{o}^w(\max_{h\geq 0}\norm{w(h)})+\gamma_\mathrm{o}^\eta(\max_{h\geq 0}\norm{\eta(h)}).
    \end{equation*}\hfill{}$\square$
\end{definition}
%
%
\end{document}